\documentclass[%
 reprint,
nofootinbib,
 amsmath,amssymb,
 aps,
prl,
floatfix,
longbibliography
]{revtex4-2}

\usepackage{todonotes}
\usepackage{graphicx}
\usepackage{dcolumn}
\newcolumntype{d}[1]{D{.}{.}{#1}}
\usepackage{bm}
\usepackage{subcaption}

\usepackage{float}
\usepackage{siunitx}
\usepackage{tabularx}
\usepackage{enumitem}

\begin{document}

\preprint{APS/123-QED}

\title{Construction of Self-Consistent Longitudinal Matches in Multi-Pass Energy Recovery Linacs}

\author{Gustavo P\'{e}rez Segurana}%
\email{g.perezsegurana@lancaster.ac.uk}
\author{Ian R. Bailey}
\affiliation{%
 Department of Physics, University of Lancaster \& Cockcroft Institute, Bailrigg, Lancaster, LA1 4YB, United Kingdom 
}%

\author{Peter H. Williams}
  \affiliation{STFC Daresbury Laboratory \& Cockcroft Institute, Warrington WA4 4AD, United Kingdom.}

\date{\today}

\begin{abstract}
Any proposal for an accelerator facility based upon a multipass Energy Recovery Linac (ERL) must possess a self-consistent match in longitudinal phase space, not just transverse phase space. We therefore present a semi-analytic method to determine self-consistent longitudinal matches in any multipass ERL. We apply this method in collider scenarios (embodying an energy spread minimising match) and FEL scenarios (embodying a compressive match), and discuss the consequences of each. As an example of the utility of the method, we prove that the choice of common or separate recirculation transport determines the feasibility of longitudinal matches in cases where disruption, such as synchrotron radiation loss, exists. We show that any high energy multipass ERL collider based upon common recirculation transport will require special care to produce a self-consistent longitudinal match, but that one based upon separate transport is readily available. Furthermore we show that any high energy multipass ERL FEL driver based upon common recirculation transport requires a larger resultant rf beam load than the one based on separate transport, favouring the separate transport designs.
\end{abstract}

\maketitle

\section{Introduction}

Energy recovery linacs (ERLs), first proposed in 1965 \cite{Tigner1965}, accelerate electron bunches to the desired target energy then decelerate the spent bunches, returning their energy to the RF system. Compared to storage rings, where electrons may circulate for hours in an equilibrium state, in ERLs the beam does not reach equilibrium. Therefore, as in traditional linacs, the beam's 6-d brightness is mainly determined by the source and may be higher than in an equivalent storage ring. A high ER efficiency then allows the average current to approach that of an equivalent storage ring. This advantageous set of properties will enable novel applications of electron accelerators in the coming decades.\par
Depending on the application, the desired characteristics of a bunch's longitudinal phase space at an interaction point can be broadly categorised into two distinct classes. If the bunch peak current is to be significantly increased upon acceleration to drive, for example, a high power FEL \cite{UK-XFEL2020ScienceCase,Benson1999} we term it a {\it compressive match}. If instead energy spread minimization is required, for example in a collider, we term it an {\it energy spread minimising match}. Of course, some situations require a partial compression, in this work we choose to explore the extremes of this continuum in order to highlight their contrasting characteristics. \par
The peak energy of an ERL provides an orthogonal categorisation of  longitudinal matches depending on whether energy loss due to synchrotron radiation is significant in comparison to the energy acceptance of the transport, or the desired dump energy. As this scales as the Lorentz factor to the fourth power~\cite{Sands1970}, realistically sized facilities can be split into those below a few GeV, and those above.\par
Finally, many different arrangements of the accelerator elements can form an ERL, however one critical characteristic of all possible topologies is whether the beam traverses the same arc\footnote{In practice, what we refer to here as ``arc" will actually comprise a spreader-arc-recombiner sequence of transport elements between linac passes or interaction regions} accelerating and decelerating, or if the beam only traverses each arc once. The former case we term {\it common transport}, the latter we term {\it separate transport}. An example of each of these is shown in Fig.~\ref{fig:Topologies cartoon}. The additional degrees of freedom available in a separate transport ERL are control of path lengths and longitudinal dispersions independently during acceleration and deceleration.\par
In this paper we explore these categories of possible ERLs and how each category exhibits a different set of possible longitudinal matches.
\section{Definitions \& Assumptions}\label{sec:DefinitionsAndAssumptions}
In setting out our general framework for constructing longitudinal matches for energy recovering systems we make the following approximations:
\begin{itemize}
    \item The quality factor of an RF system is effectively infinite. Equivalently the time taken for a bunch to transit the entire system is small compared to $Q\times T$ where $T$ is an rf period. 
    \item The bunch charge is such that the system is below any beam break-up threshold.
    \item The system is in steady state, any start-up transients have dissipated.  
\end{itemize}
As such this methodology establishes the single bunch longitudinal dynamics in steady state. The consequences of relaxing the first two conditions are explored in \cite{Setiniyaz2020,Setiniyaz2021}, where we see that the ordering of bunches, or {\it filling pattern}, affects LLRF stability and the regenerative BBU threshold. Transient effects will be explored in a subsequent paper.
\par
Each pass of the beam through an RF section represents a load. We represent this load in the complex plane as shown later in Fig.~\ref{fig:compressNoSR}. A beam on the accelerating crest is defined as $\theta=0$, with the decelerating trough being $\theta=\pi$. We can use this to illustrate the full system characteristics of a longitudinal match and determine its viability. As a first approximation, a complete energy recovery match exhibits a resultant load (vector sum of each pass) lying on the vertical axis. This corresponds to the energy transferred from the RF system to the beam during acceleration being equal to the energy deposited back from the beam to the RF cavities during deceleration. A resultant that lies exactly at the origin indicates that any off-crest acceleration is matched by corresponding off-trough deceleration\footnote{Naively one could expect that this condition guarantees that any chirp imparted to the bunch on acceleration is removed on deceleration. However this is not generally the case, we explore this point later.}. If the ERL consists of multiple RF sections, the resultant RF load of each section must lie on the vertical axis unless there is a mechanism present to transfer load between them, for example \cite{Ainsworth2016,Konoplev2020}. If energy lost to synchrotron radiation (SR) is significant, this energy balance must change. We may either reduce the energy recovery efficiency by the same amount as is lost to SR, or keep full ER but offset the dump and injector energies by the same amount. We explore the consequences of each of these choices.\par
The RF phase that the beam sees on each pass is determined by the arc path lengths and the synchronicity between the different linac sections. In a separate transport ERL, we can independently tune phases in all accelerating and decelerating passes, whereas in a common transport ERL our initial conditions and accelerating phases determine the corresponding decelerating phases.\par
The different phase choices affect the mean energy of the particles in the bunch and chirp. Depending on the system application, the fully accelerated beam may require a chirp or not. Similarly, during deceleration, as the beams relative energy spread undergoes adiabatic growth, proper setting of phases and longitudinal dispersions are required to keep the beam within the energy acceptances of the arcs.
\par
\begin{figure}[ht]
\begin{subfigure}{0.45\textwidth}
    \includegraphics[width=\textwidth]{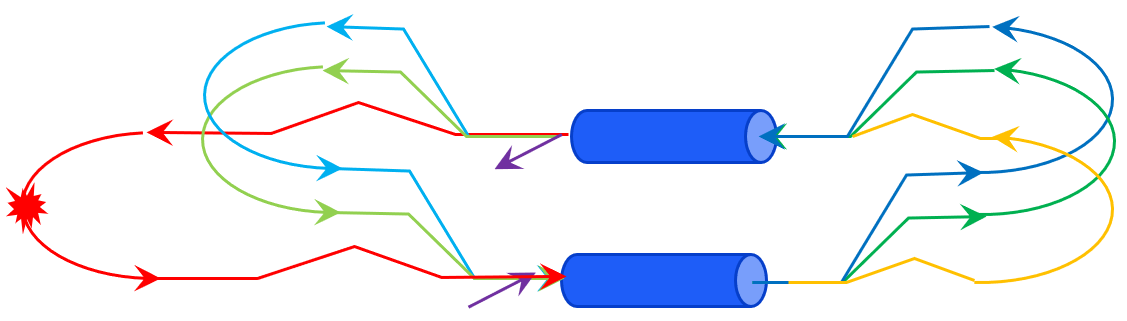}
    \caption{}\label{subfig:CommonTransportCartoon}
\end{subfigure}
\begin{subfigure}{0.45\textwidth}
    \includegraphics[width=\textwidth]{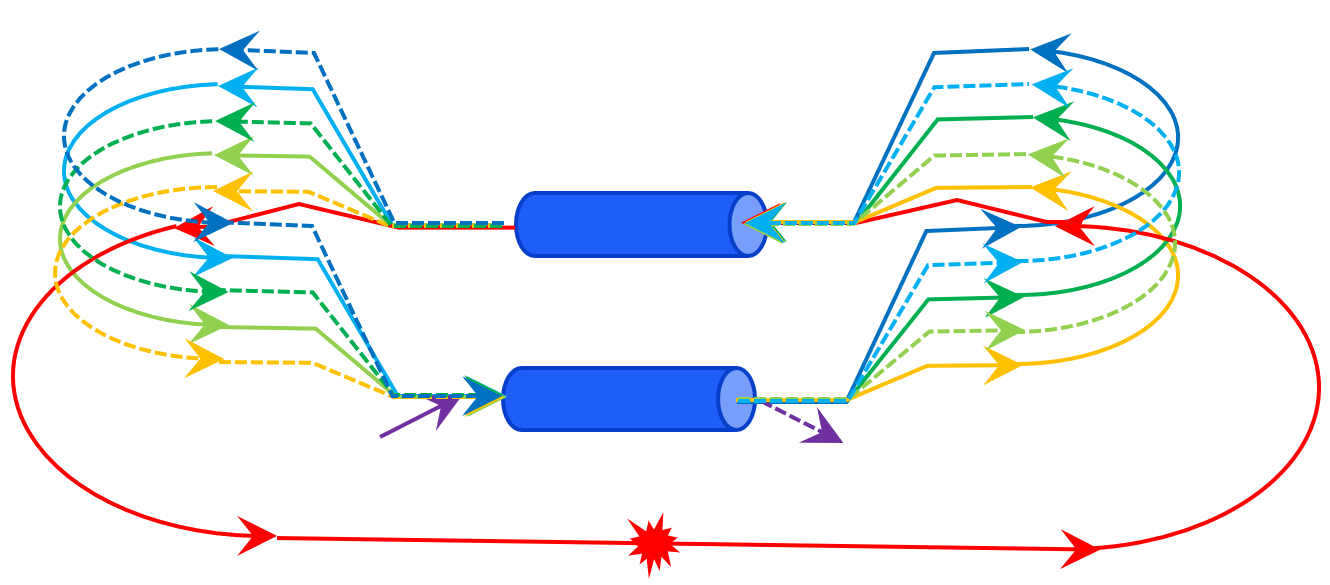}
    \caption{}\label{SeparateTransportCartoon}
\end{subfigure}
\caption{Simplest racetrack ERL configurations, blue cylinders represent the linacs, and the spiked ball represents the interaction region. (a) common transport and (b) separate transport, with solid and dashed lines indicating the the arcs traversed during acceleration and deceleration respectively.}\label{fig:Topologies cartoon}
\end{figure}

Many different configurations are possible for a common transport ERL, in this paper we focus on a racetrack configuration similar to ER@CEBAF~\cite{Bogacz2016b,Meot2016a} and PERLE~\cite{Angal-Kalinin2018}. This employs two linacs to provide higher density of accelerating sections for the same footprint as compared to a single linac such as S-DALINAC~\cite{Arnold2019}. For ease of comparison, we consider separate transport examples with topologies which match our common transport design during acceleration. However, instead of re-injecting the top energy beam into the injection linac it is re-injected into the opposing linac. In this way, accelerating and decelerating beams of the same energy traverse different arcs. Schematics of these topologies are shown in Fig.~\ref{fig:Topologies cartoon}.\par
When considering viable longitudinal matches, we favour isochronous arcs over non-isochronous ones. This is to minimize beam degradation due to collective effects that become magnified by $R_{56}$ excursions and the resulting longitudinal bunch charge distribution modulations~\cite{Tsai2017}.\par
Additionally, we must consider the implications of parasitic compressions, also know as overcompressions, where the bunch head and tail exchange places. One could expect significant degradation to occur at a parasitic compression, and it would be of particular concern during acceleration. However if the minimum bunch length during this compression is relatively large due to the presence of uncompensated RF curvature at that location, such degradation would not be significant.\par
Harmonic RF is an established technique to linearize longitudinal phase space~\cite{Zagorodnov2011}. It may also be used to top up the energy of both accelerating and decelerating beams in an ERL, in order to compensate for energy lost to synchrotron radiation. However, linearization requires deceleration of the bunch during acceleration, and acceleration during deceleration. Whereas, a compensation for SR requires always accelerating the bunch. Therefore one cannot simultaneously compensate for SR loss {\it and} linearise. Finally, the cost implications of an additional SRF system motivates the study of alternatives to correctly manipulate the longitudinal phase space. For these reasons we do not consider them in this work.
\section{Semi-analytic method}
We employ a semi-analytic method extended from that of Zagorodnov and Dohlus~\cite{Zagorodnov2011}, adding ERL operational constraints to find self-consistent longitudinal matches. An alternative strategy would be to use one dimensional longitudinal phase space particle tracking~\cite{Bane2005}. We consider this to be impractical and opaque due to the large number of discrete stages required in a multipass ERL design, which results in a solution space of very large dimension. In principle one could employ genetic algorithms or similar methods to search this space, but in doing this one loses full understanding of minimal, simplest solutions. A semi-analytic method lends itself more readily to conceptual simplicity. 
\par
The energy distribution of the initial bunch is approximated as 
\begin{equation}
    \delta_{0}(s)=
    \delta'_{0}s+
    \frac{\delta''_{0}}{2}s^{2}+
    \frac{\delta'''_{0}}{6}s^{3}\,,
\end{equation}
where $s$ is the longitudinal position of the particles in the bunch and $\delta$ is the fractional energy deviation with respect to the nominal energy. Arc elements are defined as drifts such that 
\begin{equation}
    s_{i}=
    s_{i-1}+
    (R_{56}^{(i)}\delta_{i}+
    T_{566}^{(i)}\delta_{i}^{2}+
    U_{5666}^{(i)}\delta_{i}^{3})\,,
\end{equation}
where $i$ represents the element index. RF elements are modeled as thin lenses where
\begin{equation}
    \delta_{i}= 
    \frac{(1+\delta_{i-1})E_{i-1}+\Delta E_{i}}{E_{i}}-1\,,
\end{equation}
and $E_{i}$ is the beam centroid energy at the $i^{\textrm{th}}$ stage and $\Delta E_{i}=E_{i}-E_{i-1}$. Finally, the effect of ISR is approximated by a single element such that
\begin{equation}
    U_{0}=\frac{C_{\gamma}E_{0}^{4}}{\rho_{0}}
\end{equation}
where, as introduced in \cite{Sands1970}, $U_{0}$ is the energy radiated in one revolution, by an electron bunch with nominal energy $E_{0}$, fixed radius $\rho_{0}$ and $$C_{\gamma}=\frac{4\pi}{3}\frac{r_{e}}{(m_{e}c^{2})^{3}}=\SI{8.85e-5}{m.\GeV^{-3}} ,$$ where $m_{e}$ is the electron rest mass and $r_{e}$ is the classical electron radius. Additionally, we use the inverse global compression function,
\begin{equation}
    Z_{n}=\frac{\partial s_{N}}{\partial s} .
\end{equation}
and its derivatives. Thence, we generate a system of equations describing the evolution of the longitudinal phase space of an electron bunch in an ERL analogous to eqns. (A1) and (A2) in ref.~\cite{Zagorodnov2011}.\par
\begin{figure}[!hbt]
    \centering
    \includegraphics[width=0.5\textwidth]{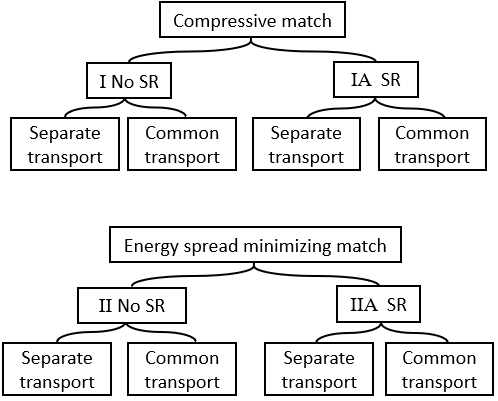}
    \caption{Classification of longitudinal matches for ERLs  whose feasability will be studied in this paper.}
    \label{fig:treeBW}
\end{figure}

Below we apply this method to a wide range of cases organized as shown in Fig.~\ref{fig:treeBW} and study their limitations as well as presenting sample solutions of each of the longitudinal matches.\par
\section{Example I: Compressive match}
A longitudinal match that increases the beam peak current from the injector to the interaction point must involve off-crest acceleration in at least one linac. By correlating the longitudinal position of the particles in the bunch with their energy, the bunch length and therefore peak current can be modulated by tuning longitudinal dispersion values. A fully compressed beam at the $i^{\textrm{th}}$ stage satisfies the condition $Z_{i}=0$.\par
\begin{figure}[ht]
 \centering
    \includegraphics[width=0.48\textwidth]{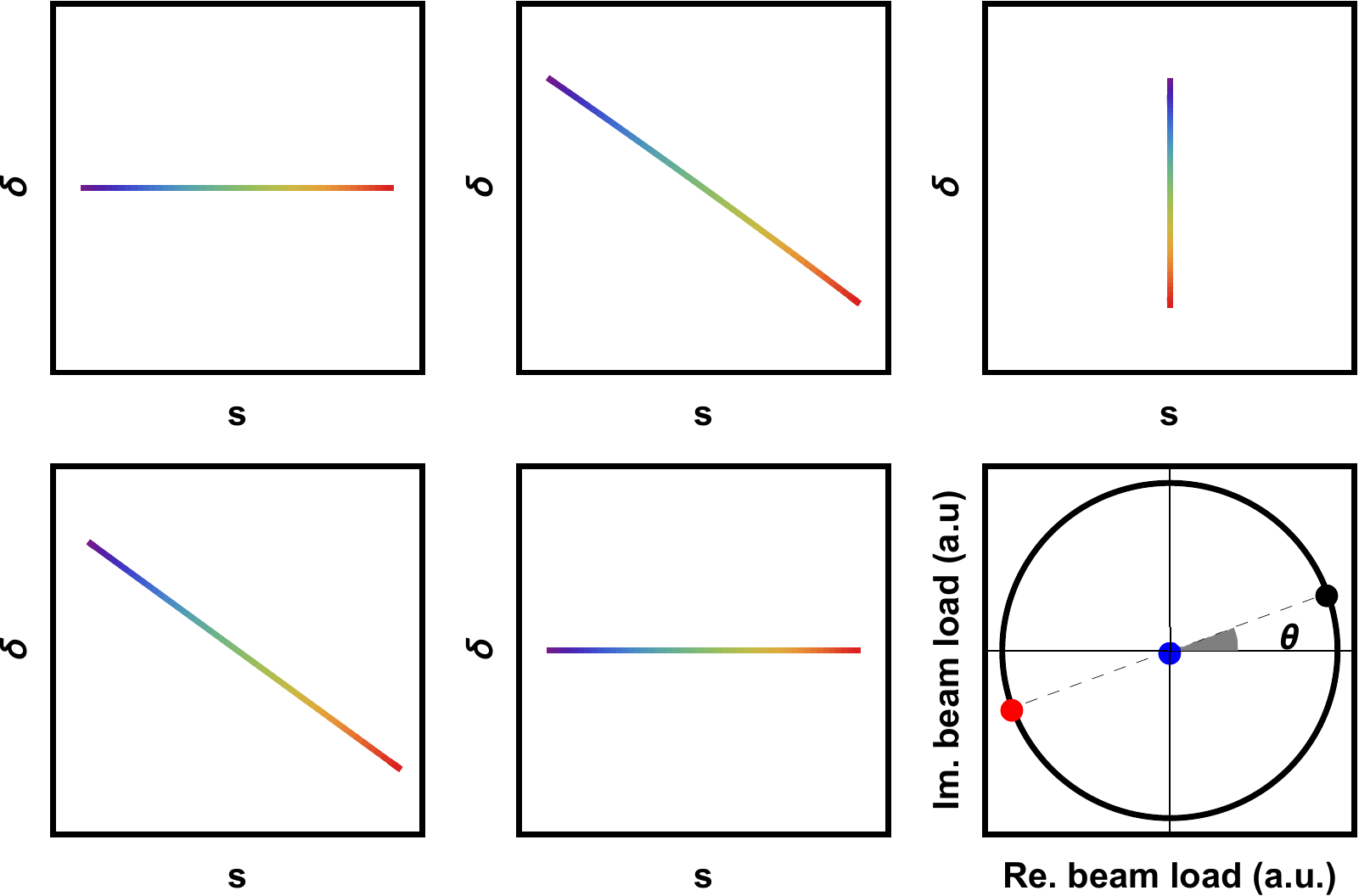}
\caption{Example I: Compressive match. Sequence of longitudinal phase space manipulations maximizing bunch current at interaction point. From top left: Initial, accelerated, compressed, decompressed and decelerated charge distributions. Bottom right shows the total rf load in the complex plane. To achieve zero chirp after deceleration we must decompress with opposite sign $R_{56}$ to that of the compression.}
\label{fig:compressNoSR}
\end{figure}

How far off crest a viable match can be is constrained by the range of $R_{56}$ available in the arcs, the energy acceptance of the arcs, and the overhead RF power available. It is advantageous to choose to accelerate on the falling side of crest as, by doing so, one utilises the natural $T_{566}$ of an arc to aid linearization~\cite{Williams2020} with a linearized bunch satisfying the condition 
\begin{equation*}
    \frac{\partial^{2}\delta}{\partial s^{2}}\Big\rvert_{s=0}=0\,.
\end{equation*}
For our first example we then select the optimal decelerating phase as that which gives zero RF load balance and compensates the beam chirp on deceleration, resulting in minimum projected energy spread at the dump.
This match is shown in Fig.~\ref{fig:compressNoSR}. The beam is accelerated $n$ times at the same rf phase, at the top energy a combination of {\it arc-like} and {\it chicane-like} sections with equal and opposite $R_{56}$ values compress and decompress the bunch. As the compression and decompression of the bunch happens at the top energy arc, this match is available in both common and separate transport configurations.

\section{Example IA: Compressive match with SR loss compensation}
The introduction of SR energy losses implies that the resulting RF load must change. We can choose to reduce the energy recovered by decelerating further off-trough than we accelerate. This change by itself however results in an overcompensation of the beam chirp, in turn this can be corrected for by modifying the decompressive $R_{56}$. By doing this we can match the accelerating and decelerating energies at a single arc, or at the dump, but not both. We are thus faced with two different scenarios depending on whether our transport is common or separate.

\subsection{Example IA with Separate Transport}
\begin{figure}[!ht]
 \centering
    \includegraphics[width=0.5\textwidth]{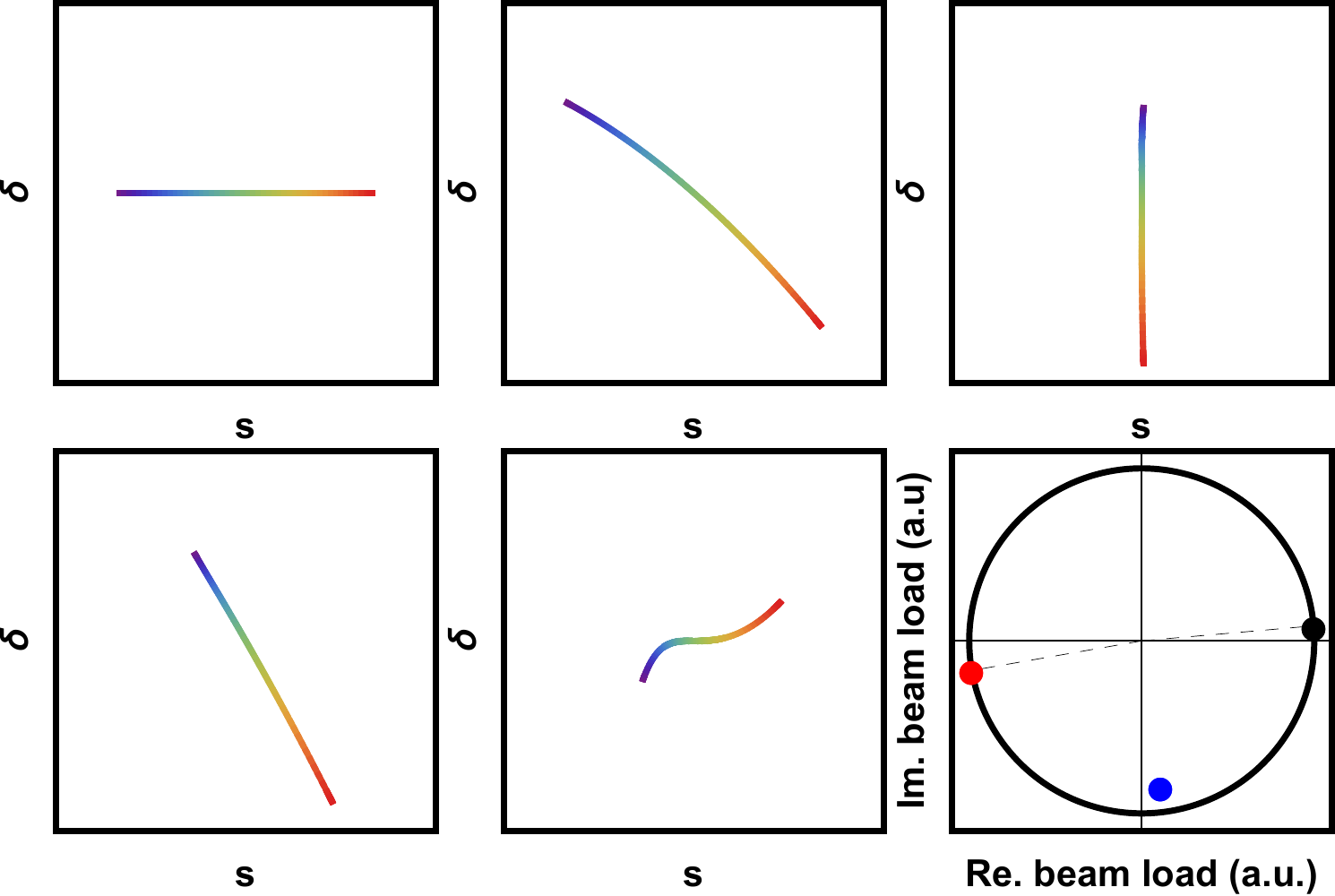}
\caption{Example IA: Compressive match with SR loss compensation. Image sequence as per Fig.~\ref{fig:compressNoSR}. Choosing decompressing $R_{56}$ of equal and opposite sign to compressing now results in finite residual chirp as we must move decelerating phase further off-trough to account for SR energy loss (resulting beam load x 10 for clarity).
}
\label{fig:compressSR}
\end{figure}

In separate transport we retain independent control over all steps as there is no need to fit both accelerating and decelerating beams in a single arc energy acceptance. It is also possible to handle larger disruptions at the interaction point, such as increased energy spread due to an FEL~\cite{Douglas2001GERBAL,Piot2003,Nergiz2021}. The independent control of longitudinal dispersions enables linearization during acceleration and deceleration as well as bunch length control. This is illustrated in Fig.~\ref{fig:compressSR}

\subsection{Example IA with Common Transport}
\begin{figure}
 \centering
    \includegraphics[width=0.5\textwidth]{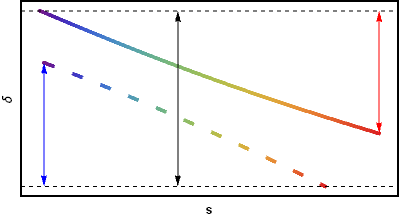}
\caption{Longitudinal phase spaces of accelerating (solid) and decelerating (dashed) bunches in an intermediate arc. The required energy acceptance for common transport corresponds to the height of the black arrow. The required energy acceptance in the corresponding separate transport configuration corresponds to only the height of the red and blue arrows for the accelerating and decelerating arcs respectively.}
\label{fig:compressSRcommonEnergyAcceptance}
\end{figure}

A comparison of the required energy acceptance between compressive longitudinal matches in common and separate transport is shown in Fig.~\ref{fig:compressSRcommonEnergyAcceptance}. As the energy lost to SR increases, the difference between the average energy of accelerating and decelerating beams will also increase. First, limiting how far off-crest the accelerator can be run, and ultimately requiring unfeasibly large energy acceptance. Additionally, the path length symmetry between acceleration and deceleration passes does not match the energy asymmetry. Therefore, if we choose to match the energy compensation we cannot match the chirp compensation.
One proposed method that has been suggested to remedy this include additional ``SR compensating linacs". However as mentioned previously these must operate at a higher even harmonic to add energy to both accelerating and decelerating beams. Linearization, requiring odd harmonics, is not possible in this scenario thereby precluding a self-consistent longitudinal match.

\section{Example II: Energy spread minimization}
\begin{figure}[!ht]
\begin{subfigure}{0.48\textwidth}
    \centering
    \includegraphics[width=\textwidth]{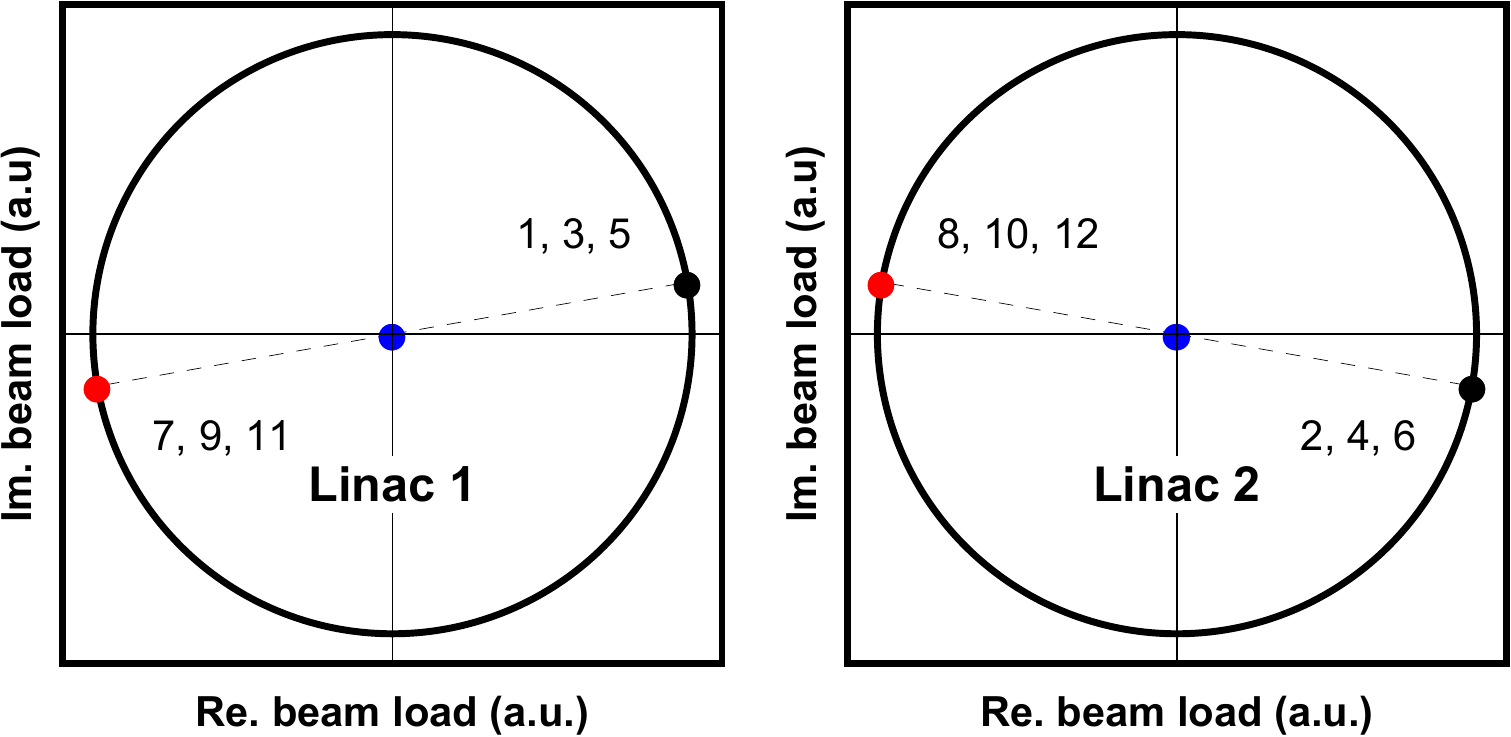}
    \vspace{-1em}
    \caption{}
    \label{fig:PERLEphasechoicesA}
\end{subfigure}
\\\vspace{1em}
\begin{subfigure}{0.48\textwidth}
    \centering
    \includegraphics[width=\textwidth]{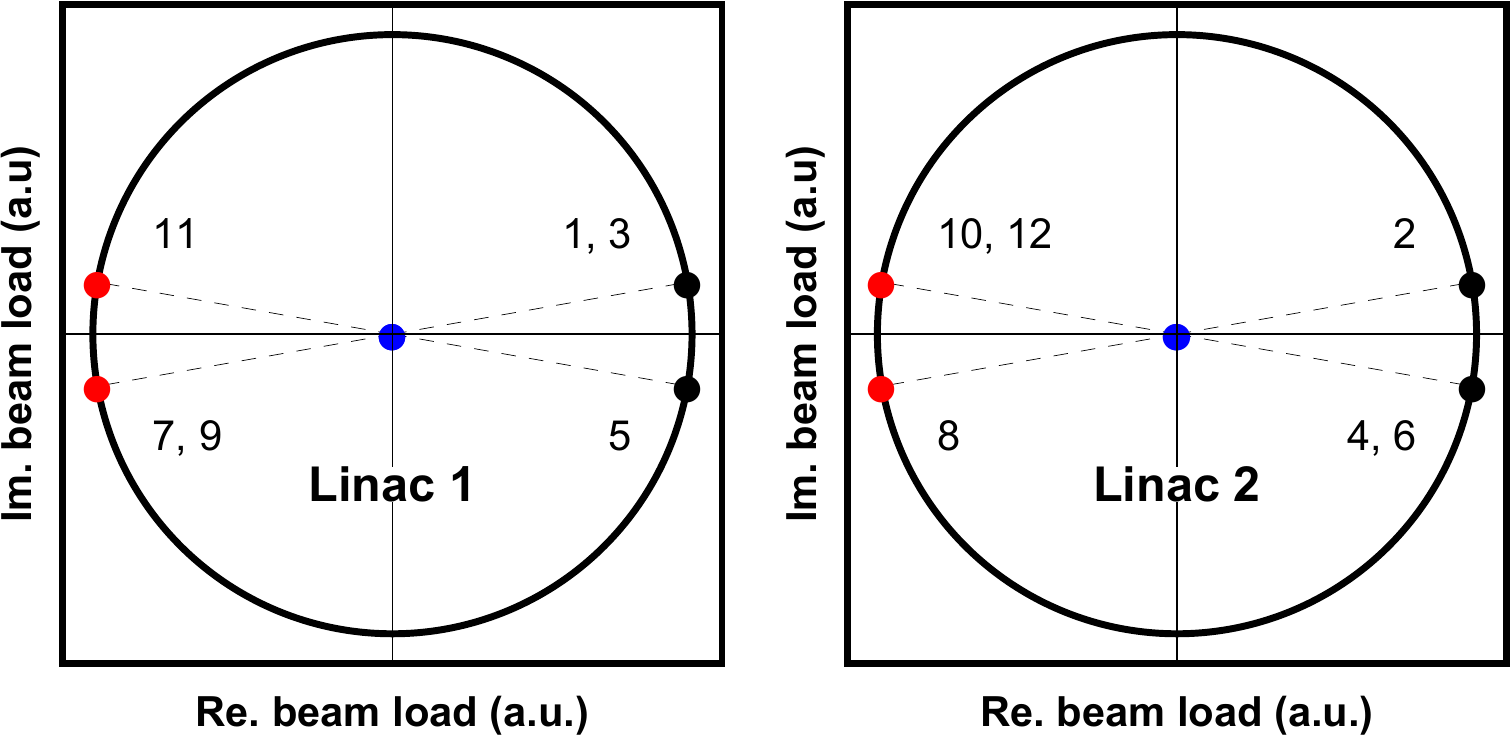}
    \vspace{-1em}
    \caption{}
    \label{fig:PERLEphasechoicesB}
\end{subfigure}
\\\vspace{1em}
\begin{subfigure}{0.48\textwidth}
    \centering
    \includegraphics[width=\textwidth]{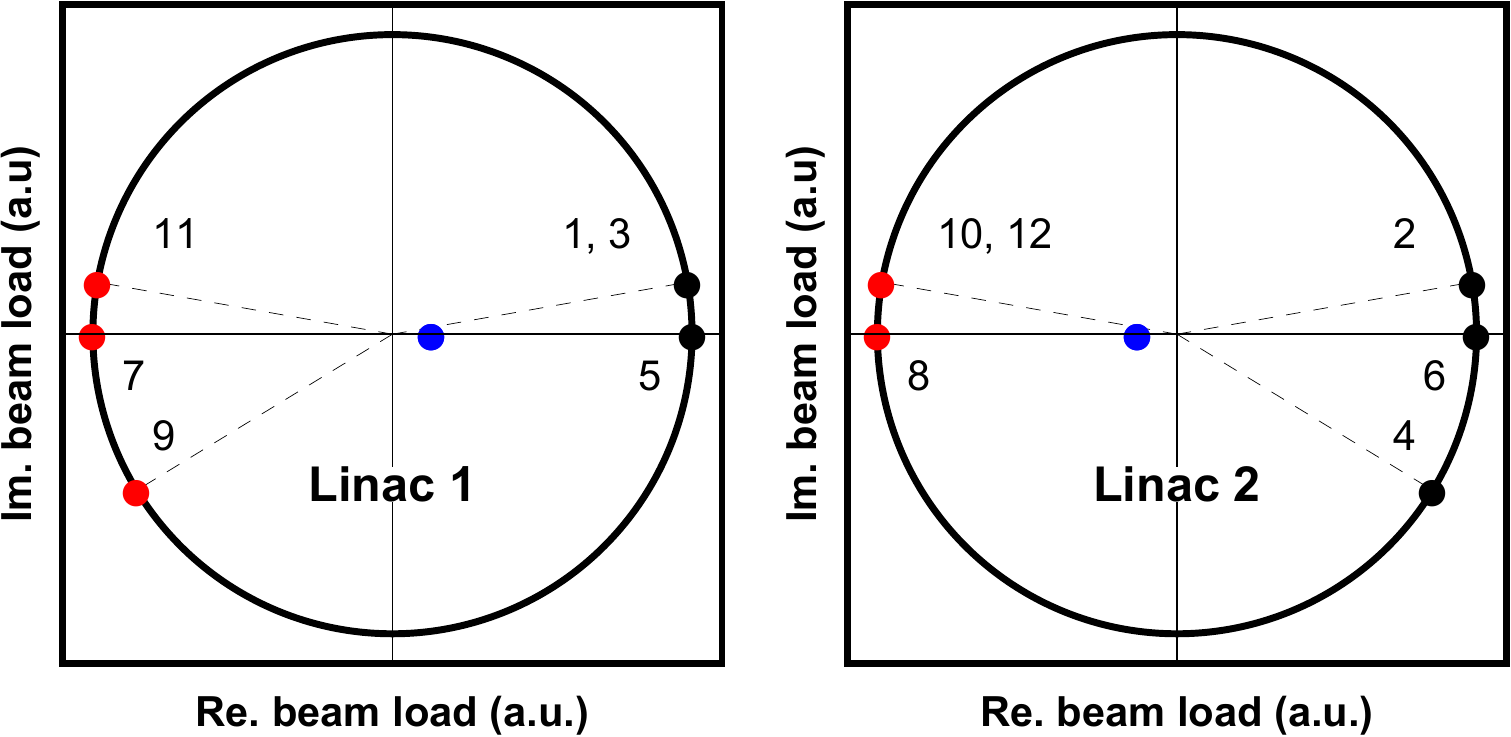}
    \vspace{-1em}
    \caption{}
    \label{fig:PERLEphasechoicesC}
\end{subfigure}
    \caption{RF beam load plots for different phase configurations in common transport longitudinal matches that minimize beam energy spread. Applicable to e.g. PERLE. RF phase choices during acceleration (black), deceleration (red), and resulting beam load (blue) of each linac independently. Number labels indicate the ordering of the RF passes.}
    \label{fig:PERLEphasechoices}
\end{figure}

\begin{figure}[!hbt]
    \centering
    \includegraphics[width=0.5\textwidth]{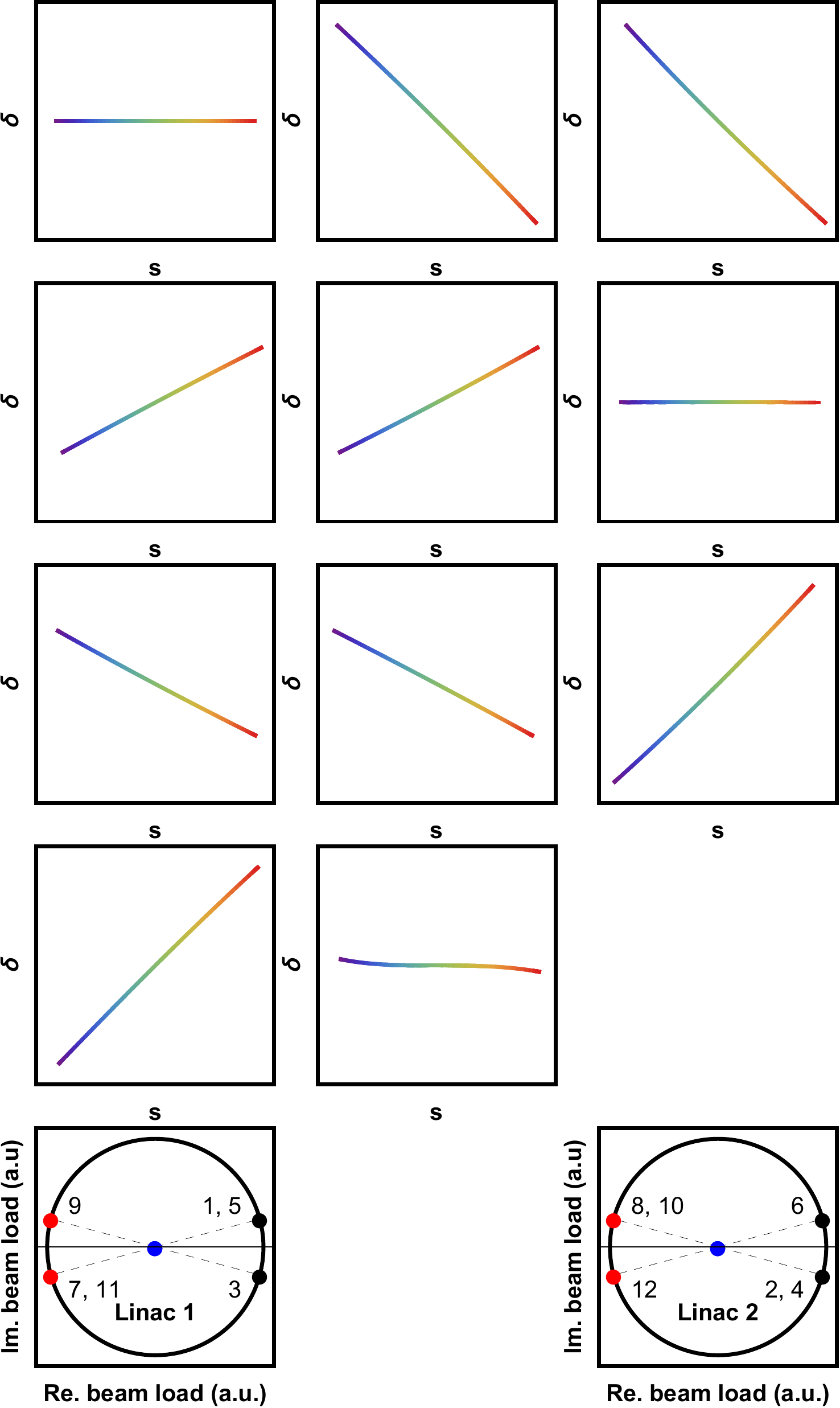}
    \caption{Energy spread minimization with common transport. Sequence of longitudinal phase space manipulations with two linearizing arcs with opposite sign of $T_{566}$. Beam load plots for the two linacs are shown underneath with labels indicating the order of each of the rf passes.}
    \label{fig:PERLELPS}
\end{figure}

A longitudinal match that delivers to the interaction point a bunch with minimal energy spread can be obtained to first order by accelerating on crest. In this case the magnitude of the absolute energy spread will be determined by the RF curvature imprinted onto the bunch during acceleration. This can correspond to several times the slice energy spread as a function of the bunch length, energy gain between injected bunch and top energy, and rf frequency.
In order to linearize the longitudinal phase space at the interaction point without using harmonic cavities, the bunch must be accelerated off-crest and the arc $T_{566}$ adequately set. For the final acceleration one must switch to the opposite side of crest in order to cancel the chirp prior to the top energy arc and interaction region, resulting in a flat bunch in longitudinal phase space. There are then three different phase setups possible that satisfy these conditions, illustrated in Fig.~\ref{fig:PERLEphasechoices}:
\begin{enumerate}[label=(\alph*)]
    \item The simplest solution runs the first linac ahead of crest and the second linac equally far behind crest, Fig.~\ref{fig:PERLEphasechoicesA}. The bunch is thus chirped into all odd arcs and dechirped into all even arcs. The $T_{566}$ of the odd arcs can be tuned to minimize the projected energy spread at the interaction point. As the beam energy increases, the beam chirp in higher acceleration passes decreases adiabatically, reducing the effect of our linearizing $T_{566}$ in arcs 3 and above.
    \item Arc pathlengths can be set such that the first half of accelerating passes are on the same phase and the second half on the opposite side of crest. This set of phases enables sharing the linearizing effort between all the arcs, with decreasing impact of the second half of the arcs, Fig.~\ref{fig:PERLEphasechoicesB}. 
    \item We may retune the previous solution such that the beam chirp from the first half of acceleration passes is completely compensated by the following pass, and the remaining accelerating passes are made on crest, Fig.~\ref{fig:PERLEphasechoicesC}. This both maximizes the effect of our linearization in the low energy arcs, and minimizes the beam energy spread in the higher energy arcs. This results in an overall reduction of sensitivity to chromatic effects. However there is not a constant energy gain between consecutive arcs and there is an energy imbalance between the two linacs, i.e. one linac recovers more energy than it uses to accelerate the beam, and the other does the opposite. In this instance a twin axis linac is required with an efficient transfer of rf power between the cavities~\cite{Konoplev2020,Konoplev}.
\end{enumerate}
All these configurations rely on chirping the beam such that during transport, the natural $T_{566}$ of the arcs has a linearizing effect. However, this significantly lengthens the low energy tail resulting in an overall longer bunch. This then covers more degrees of the RF waveform during deceleration resulting in larger energy spreads at the dump and potentially compromising the energy recovery. Instead, it is possible to set the phases such that the beam chirp has different signs as it travels through at least two of the arcs. By exchanging the role of the low energy tail in the two linearizing arcs we can keep bunch length under control. This however requires changing the sign of the $T_{566}$ of one of the arcs. An example of a suitable match is shown in Fig.~\ref{fig:PERLELPS}.\par
Overall, if all arcs are kept first order isochronous, all viable phase choices that minimize the beam energy spread at the interaction point have RF load vector sums lying on the horizontal axis. Additionally, the path length shift into the decelerating passes is such that the phases are symmetric about the vertical axis and so the rf load vector sum lies at the origin.\par
\begin{figure}[!ht]
\begin{subfigure}{0.45\textwidth}
    \centering
    \includegraphics[width=\textwidth]{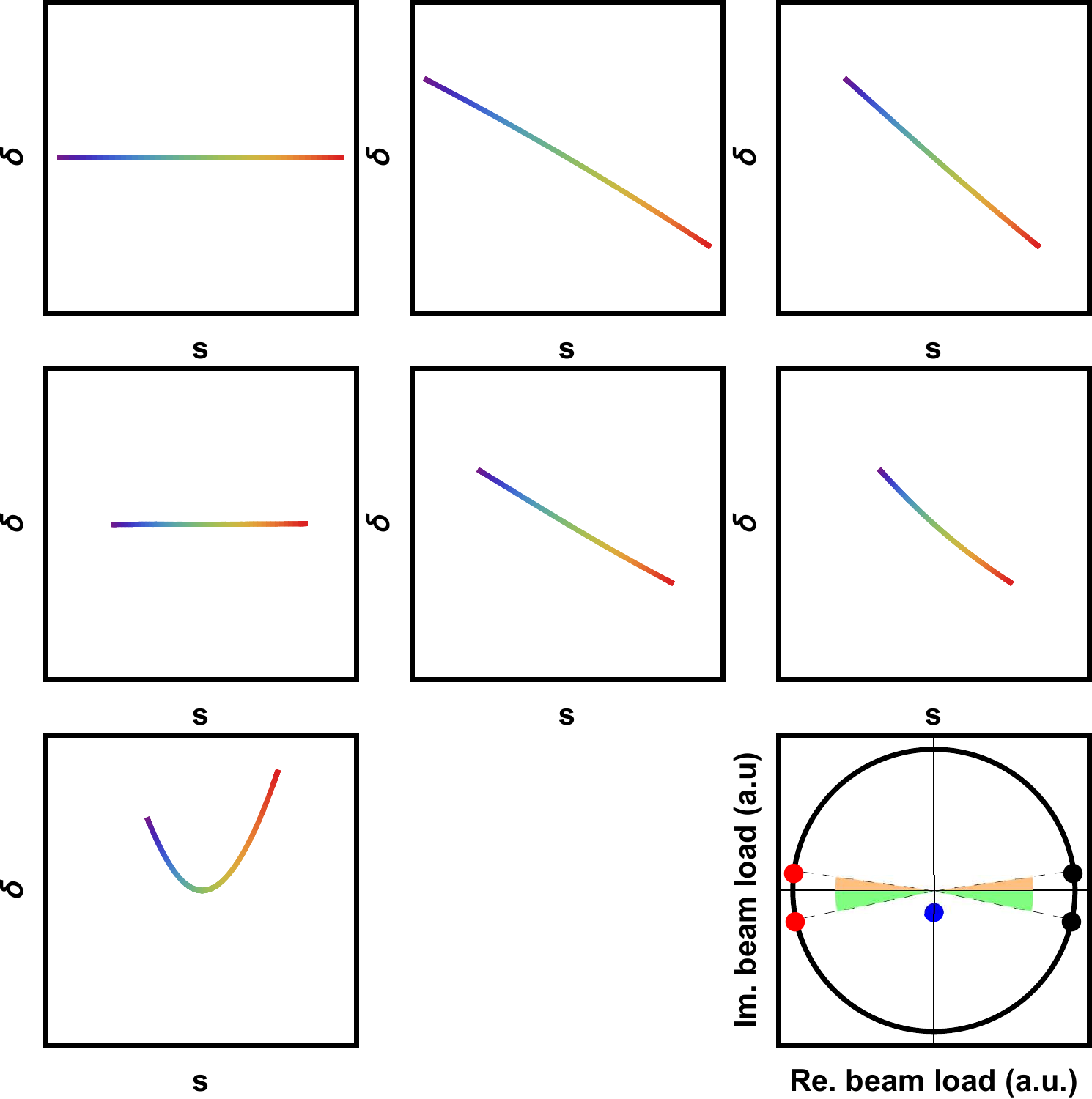}
    \vspace{-1.5 em}
    \caption{}
    \label{fig:NonIsoCommonLinearize}
\end{subfigure}
\\\vspace{1 em}
\begin{subfigure}{0.45\textwidth}
    \centering
    \includegraphics[width=\textwidth]{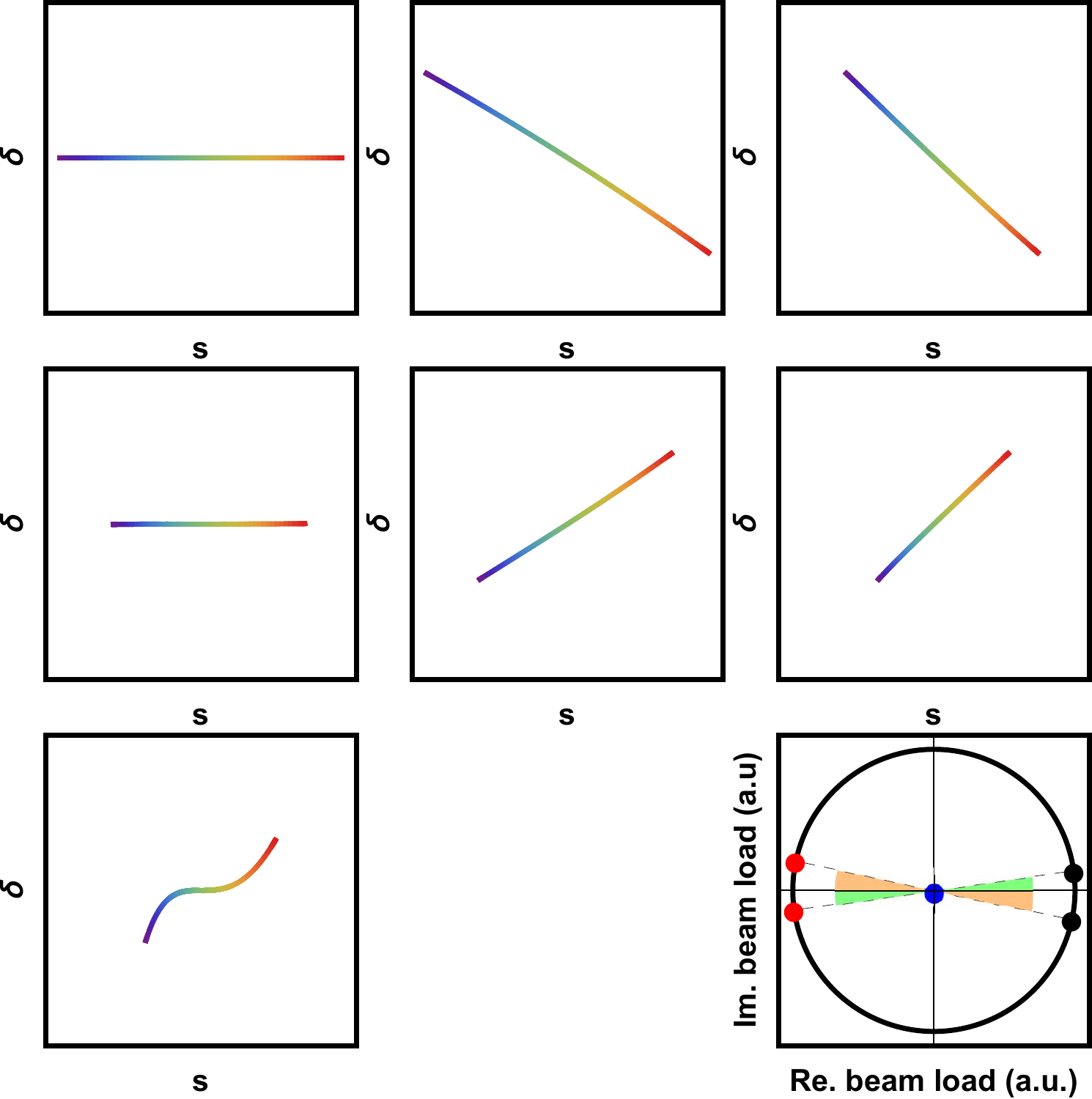}
    \vspace{-1.5 em}
    \caption{}
    \label{fig:NonIsoSeparateLinearize}
\end{subfigure}
\caption{Energy spread minimizing matches with non-isochronous intermediate arcs. RF beam load plot shows rf phase choices during acceleration (black), deceleration (red), and resulting beam load (blue). (a) Common transport, (b) separate transport. Different angle highlights in the rf load plots correspond to different magnitudes off-crest or off-trough.}
\label{fig:NonIsoLinearize}
\end{figure}

Energy spread minimizing matches are not strictly required to have a zero $R_{56}$, the beam can have a zero chirp as it reaches the IP without a purely real resulting accelerating RF load.
This series of longitudinal phase space manipulations are however limited by the requirement of a bunch at the dump to fit within the energy acceptance. Sample viable configurations are shown in Fig.~\ref{fig:NonIsoLinearize}, with Fig.~\ref{fig:NonIsoCommonLinearize} as a common transport example and Fig.~\ref{fig:NonIsoSeparateLinearize} as a separate transport example. The common transport solution shows a longitudinal phase space at the dump with the characteristic shape of the decelerating RF curvature. This is because the intermediate arc is used to linearize towards the interaction point and therefore is not a free parameter to linearize the bunch towards the dump. On the contrary, the separate transport solution shows only a third-order dependence of $\delta$ on $s$ at the dump since accelerating and decelerating arcs can be tuned to linearize the bunch at the IP and at the dump. Finally, comparing the rf loads in both cases, the common transport solution has a non-zero resultant rf load. It can be made zero in the separate transport case thanks to the independent control of the arc path lengths accelerating and decelerating.

\section{Example IIA: Energy Spread Minimization with SR loss compensation}
\subsection{Example IIA with Common Transport}
Proposed facilities above a few GeV cannot neglect SR energy losses. If these losses are small, the phase schemes above can be adapted by changing the path length of the top energy arc. However, this results in an overall chirp in the bunch as it reaches the dump. Therefore, the limits of this strategy are defined by the necessary decelerating phase shift to compensate for the losses, and by the energy acceptance of the arcs and dump.\par
\begin{figure}[!htb]
    \centering
    \includegraphics[width=0.48\textwidth]{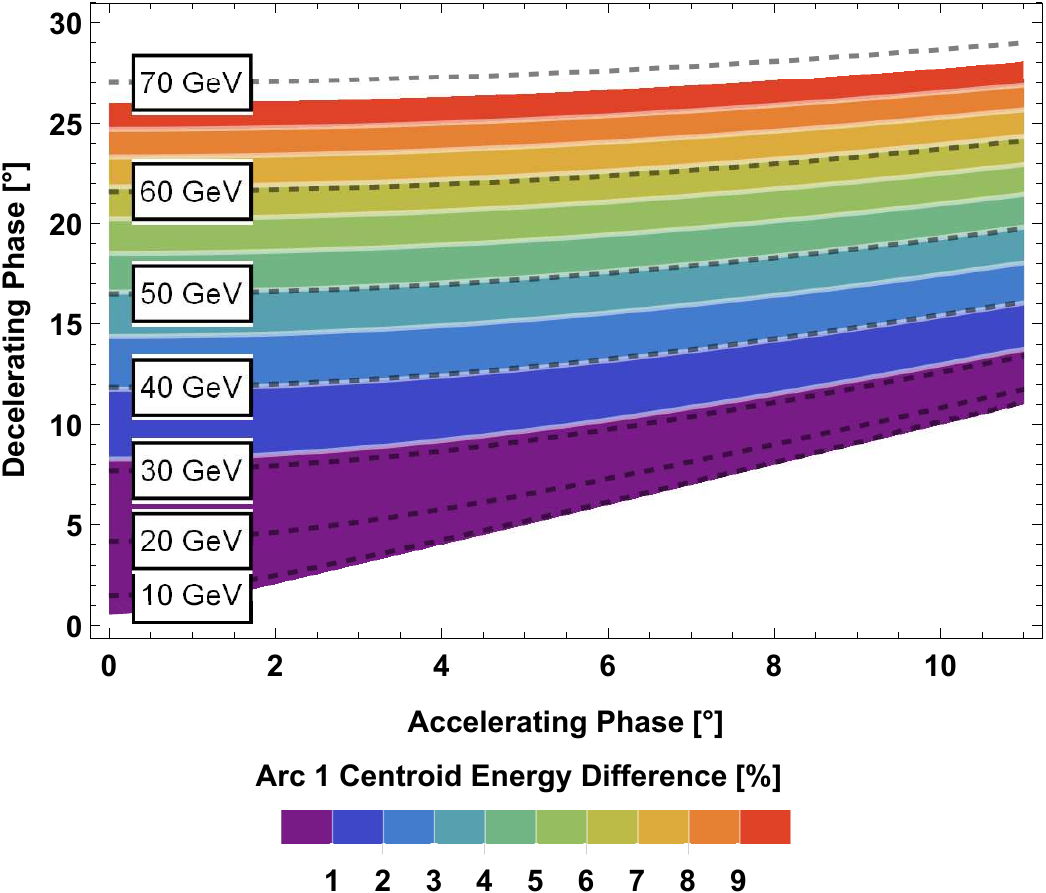}
    \caption{Arc 1 centroid energy difference between accelerating and decelerating beams for a range of accelerating and decelerating phases. Black dashed contour lines represent peak energy in $\SI{}{GeV}$.}
    \label{fig:LHeCArc1Acceptance}
\end{figure}

For non-negligible energy loss, tuning the top energy arc path length can only match the decelerating energy at a single stage. This results in differences in centroid energies at all other stages, requiring very large energy acceptances in these arcs even before taking into account the bunch energy spread. Figure~\ref{fig:LHeCArc1Acceptance} shows the energy acceptance necessary in arc 1 of a 3-turn (accelerating and 3-turn decelerating) common transport ERL for a range of peak energies with energy losses corresponding to 180 degree arcs containing dipoles with a geometric radius of $\SI{336}{m}$, similar to those proposed for LHeC \cite{LHeCandFCC-heStudyGroup2020}.\par

Alternatively, all arc path lengths may be used to set the RF phases such that the difference between centroid energies is minimized, reducing the minimum arc energy acceptance~\cite{Koscica2019}. Example results of such a minimization are shown in Table~\ref{tab:LHeCNminimize}. We have shown previously that longitudinal matches that minimize energy spread at the interaction point are possible even if the the resulting accelerating RF load has a non-zero but small imaginary component. The RF phase choices necessary in this case result in very far off crest phases, in turn resulting in chirped beams with very large energy spread, a magnification of the effect shown in Fig.~\ref{fig:compressSRcommonEnergyAcceptance}.

\begin{table}[!ht]
    \begin{tabular}{ccc}
        Parameter & Value & Units \\ \hline\hline
        $\phi_{1,1}$ & $0.0$ & $^\circ$ \\
        $\phi_{1,2}$ & $-13.5103$ & $^\circ$\\
        $\Delta\phi_{1}$ & $0.0$ & $^\circ$ \\
        $\Delta\phi_{2}$ & $-0.755549$ &$^\circ$\\
        $\Delta\phi_{3}$ & $44.2088$ & $^\circ$\\
        $\Delta\phi_{4}$ & $-13.6424$& $^\circ$ \\
        $\Delta\phi_{5}$ & $13.5607$ & $^\circ$\\
        $\Delta\phi_{6}$ & $100.057$ & $^\circ$\\
        $\delta_{0,\textrm{min}}$ & $0.819755$ & \% \\
        $\delta_{1,\textrm{min}}$ & $0.593161$ & \% \\
        $\delta_{2,\textrm{min}}$ & $0.768878$ & \% \\
        $\delta_{3,\textrm{min}}$ & $0.693516$ & \% \\
        $\delta_{4,\textrm{min}}$ & $0.647318$ & \% \\
        $\delta_{5,\textrm{min}}$ & $0.822313$ & \% \\
    \end{tabular}
    \caption{Results of numerical optimization of arc path lengths and initial RF phases that minimize the difference in relative momentum between accelerating and decelerating beams traversing the same arcs of a LHeC-like machine. $\phi_{1,1}$ and $\phi_{1,2}$ are the initial phases of the linacs,  $\Delta\phi_{1}$ through  $\Delta\phi_{6}$ are the phase changes between rf passes at each of the arcs and  $\delta_{0,\textrm{min}}$ through $\delta_{5,\textrm{min}}$ are the fractional energy acceptances necessary to accommodate both the accelerating and decelerating beams from the Injector/Dump to arc 5.}
    \label{tab:LHeCNminimize}
\end{table}
\subsection{Example IIA with Separate Transport}
\begin{figure*}[!htb]
    \centering
    \includegraphics[width=\textwidth]{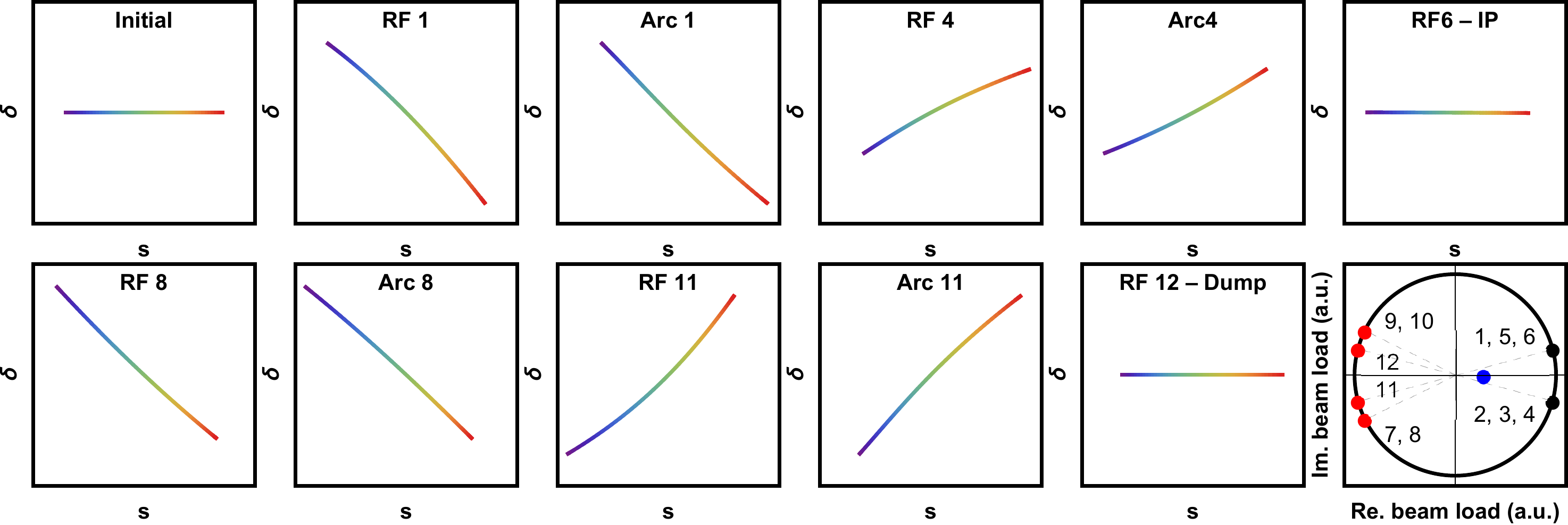}
    \caption{Energy spread minimizing match with SR loss compensation using separate transport. Shown longitudinal phase spaces correspond to the example beam at the exit of the element specified. Applicable to e.g. LHeC. 
    RF beam load plot shows phase choices during acceleration (black), deceleration (red), and resultant (blue).}
    \label{fig:LHeCSeparate}
\end{figure*}

A separate transport solution is readily available since accelerating and decelerating beams need not share the same centroid energy. Such a solution is shown in Fig.~\ref{fig:LHeCSeparate}.
In this example the bunch is accelerated off-crest in a similar fashion as the example of Fig.~\ref{fig:PERLEphasechoicesB} with linearizations in arc 1 and arc 4. Then, the first 4 decelerating passes are further off-crest than their accelerating counterparts to compensate for the energy losses whilst keeping a purely real resultant beam load in both linacs. The final two decelerating phases are equally as far off-crest as the accelerating passes to control the energy spread in the low energy arcs. Linearizations during deceleration occur in arc 8 and arc 10  (energy levels 4 and 1).

\section{Bunch length control through alternate sign linearization}
One method of increasing luminosity is to increase bunch charge by allowing longer bunches from an injector. Pre-compression of such bunches in an injector chicane or equivalent may not be optimal due to emittance degradation through collective effects.\par
\begin{figure}[!ht]
\begin{subfigure}{0.4 \textwidth}
    \includegraphics[width=\textwidth]{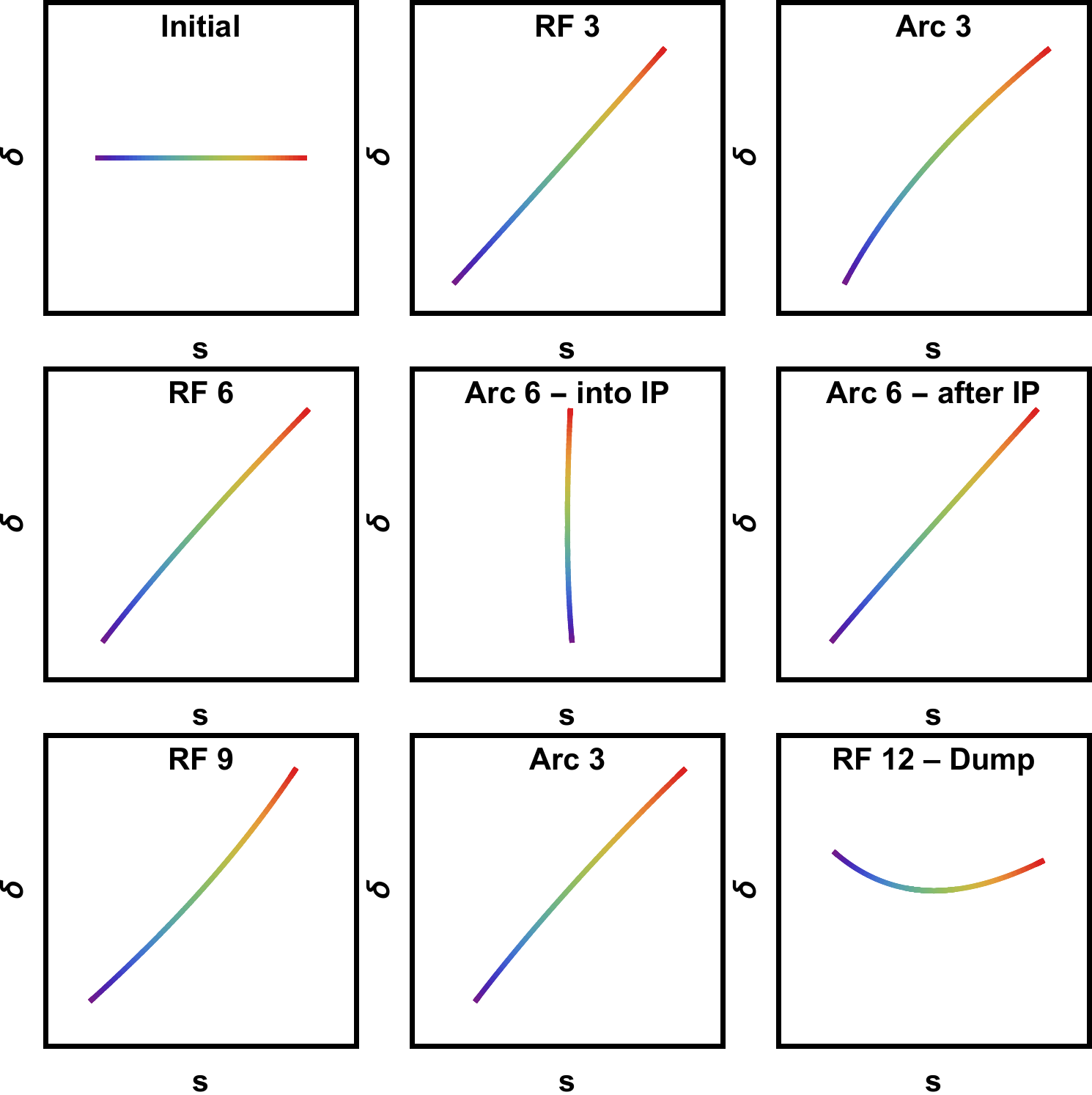}
    \vspace{-1.2em}
    \caption{}\label{subfig:BunchLengthControla}
\end{subfigure}
\\\vspace{1.2 em}
\begin{subfigure}{0.4 \textwidth}
    \includegraphics[width=\textwidth]{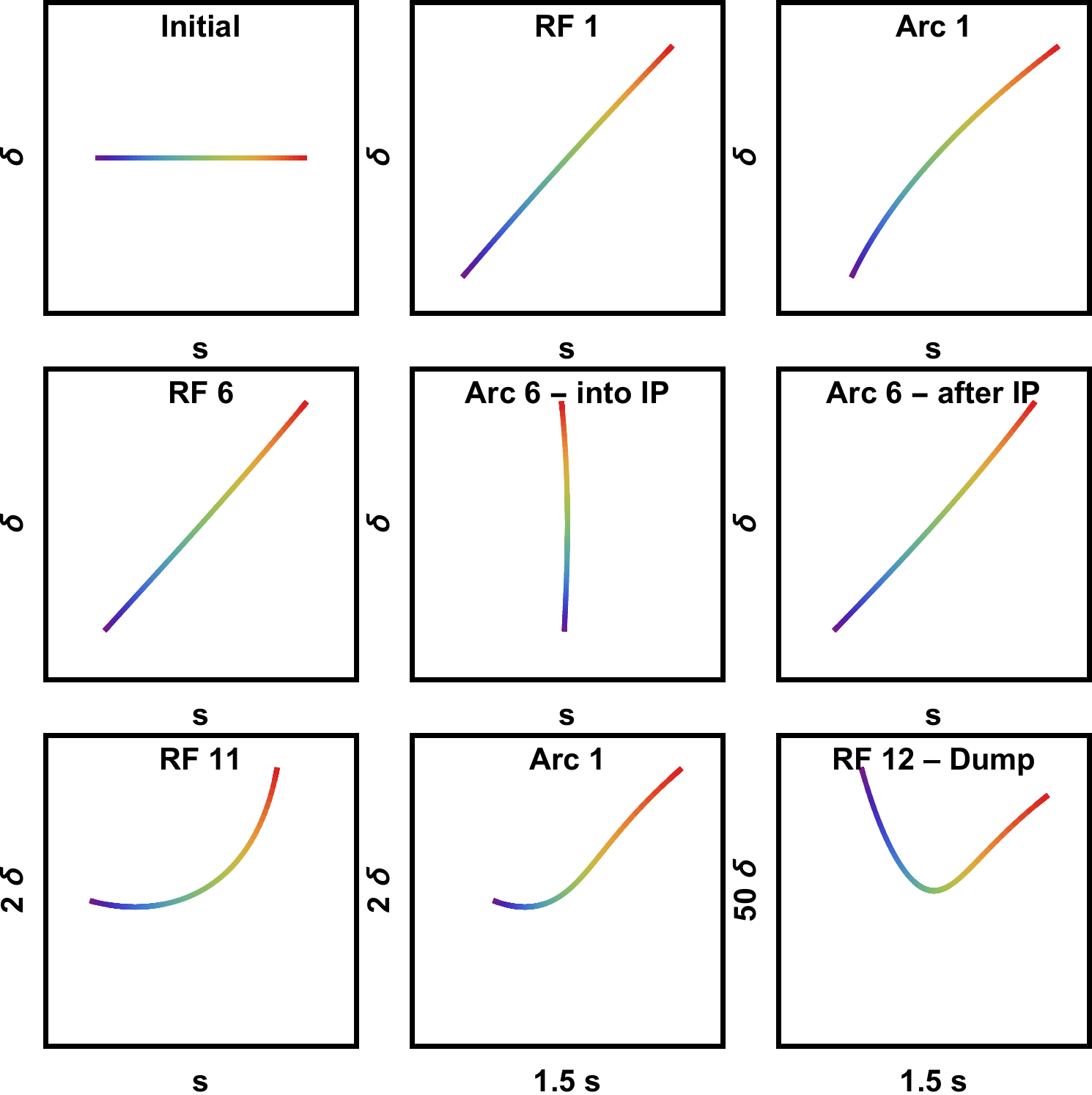}
    \caption{}\label{subfig:BunchLengthControlb}
\end{subfigure}
\caption{Sequences of compressive common transport longitudinal matches with different choice of linearizing arcs. Longitudinal phase spaces at the exit of the specified elements. The remaining arcs have a non-zero $T_{566}$ of their natural sign. (a) uses arc 3 to linearize, (b) uses arc 1 to linearize. Note the change of scale in the later stages of (b).}
\label{fig:BunchLengthControl}
\end{figure}

Linearization of the longitudinal phase space by controlling second order longitudinal dispersions during acceleration results in a compression of the high energy tail and elongation of the low energy tail. Depending on the strength of the linearization required, the resulting bunch elongation may not be tolerable. This bunch elongation can be controlled by splitting the linearization process into several steps and utilising arcs with $T_{566}$ values with opposite signs. This can be achieved by pre-linearizing the bunch in an injection chicane before entering the main ERL loop, or by setting the accelerating RF phases such that in the first accelerating pass the beam chirp is of the opposite sign to that of the fully accelerated beam. This change in rf phase choices will result in an increment in the RF load for non SR compensating common transport configurations, and a larger centroid energy mismatch in the intermediate arcs of a SR compensating common transport accelerator. Alternatively, as shown in Fig.~\ref{fig:BunchLengthControl} careful selection of a single linearizing arc can result in successful matches if the natural, non-zero $T_{566}$ of the rest of the arcs is taken into account. In Fig.~\ref{subfig:BunchLengthControla} we demonstrate a successful match by arranging a balance between a natural over-linearizing $T_{566}$ of the arcs and the chosen arc 3 which is anti-linearizing. Conversely, in Fig.~\ref{subfig:BunchLengthControlb} we show the negative consequences if arc 1 is chosen to control the linearization: As we decelerate through arcs 5 to 2 the low energy tail is elongated and the high energy tail compressed, changing the profile of the curvature imprinted onto the bunch resulting with an energy spread for the fully-decelerated bunch which is much too large.
\section{Strategies to mitigate common transport limitations}
Employing the same arcs during acceleration and deceleration limits the control over path lengths and longitudinal dispersions, whilst sharing the same momentum acceptance. However, a more complex design of the transport can mitigate this.\par
\begin{figure}[!ht]
    \centering
    \includegraphics[width=0.48\textwidth]{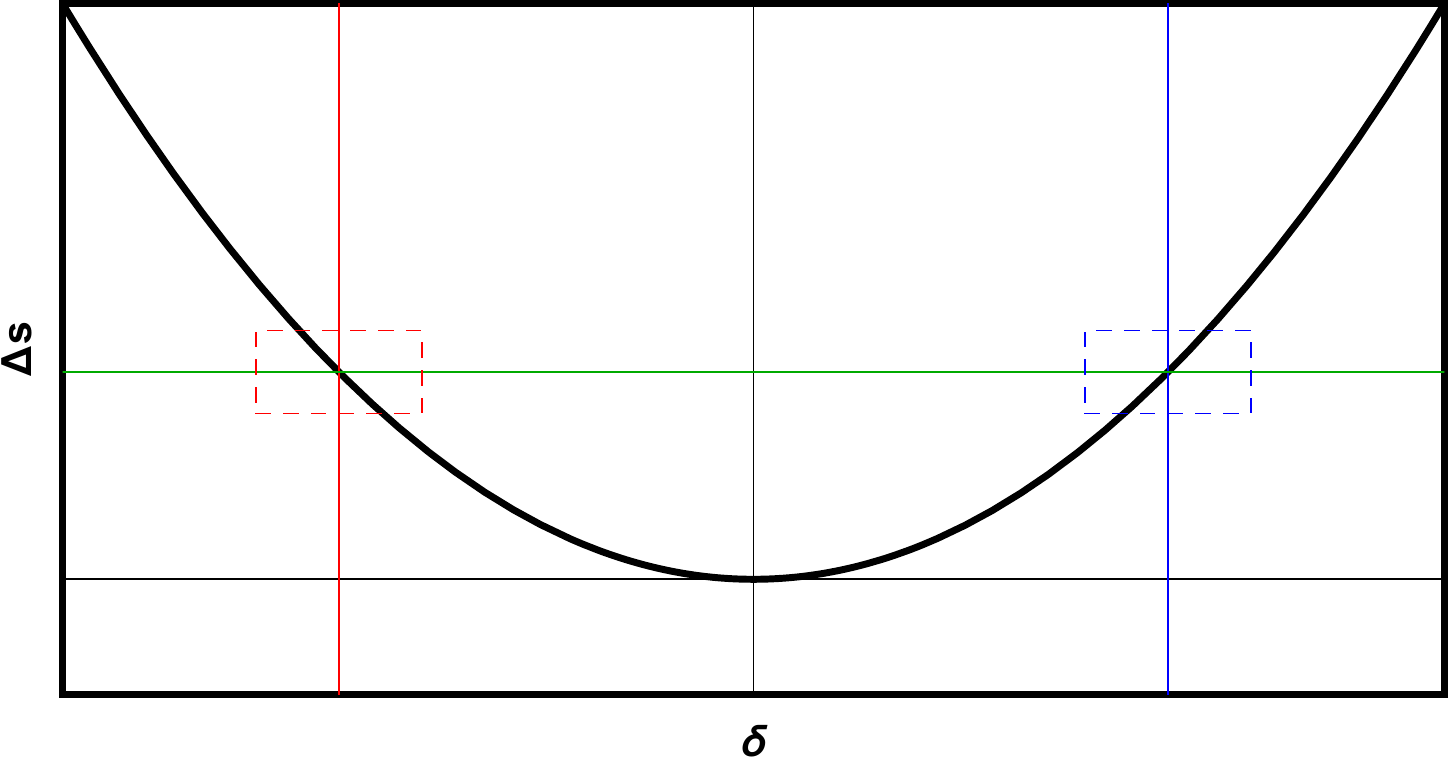}
    \caption{Arc path length as a function of relative momentum deviation with only second order longitudinal dispersion non-zero. Red and blue displaced axes highlight the path length dependence on momentum for off-momentum beams with an effective non-zero $R_{56}$.}
    \label{fig:t566compressdecompress}
\end{figure}

In a common transport configuration, if the top energy arc cannot reach the necessary $R_{56}$ values, a large $T_{566}$ in the second-to-top arc may be set to compress accelerating bunches and decompress decelerating bunches as shown in Fig.~\ref{fig:t566compressdecompress}. This is thanks to the difference in centroid energies between the accelerating and decelerating beams which correspond to the sum of SR losses and any energy lost at interaction. However, in doing this we must transport a compressed bunch for longer, risking collective effects degrading the bunch prior to interaction. Additionally, in order to effectively transport the beam in this arc, the arc must have good chromatic behaviour over the whole range of the energy acceptance including zeroing higher order transverse dispersions and chromatic amplitudes~\cite{Lindstrm2016,Williams2020}.\par
\begin{figure}[!ht]
    \centering
    \includegraphics[width=0.48\textwidth]{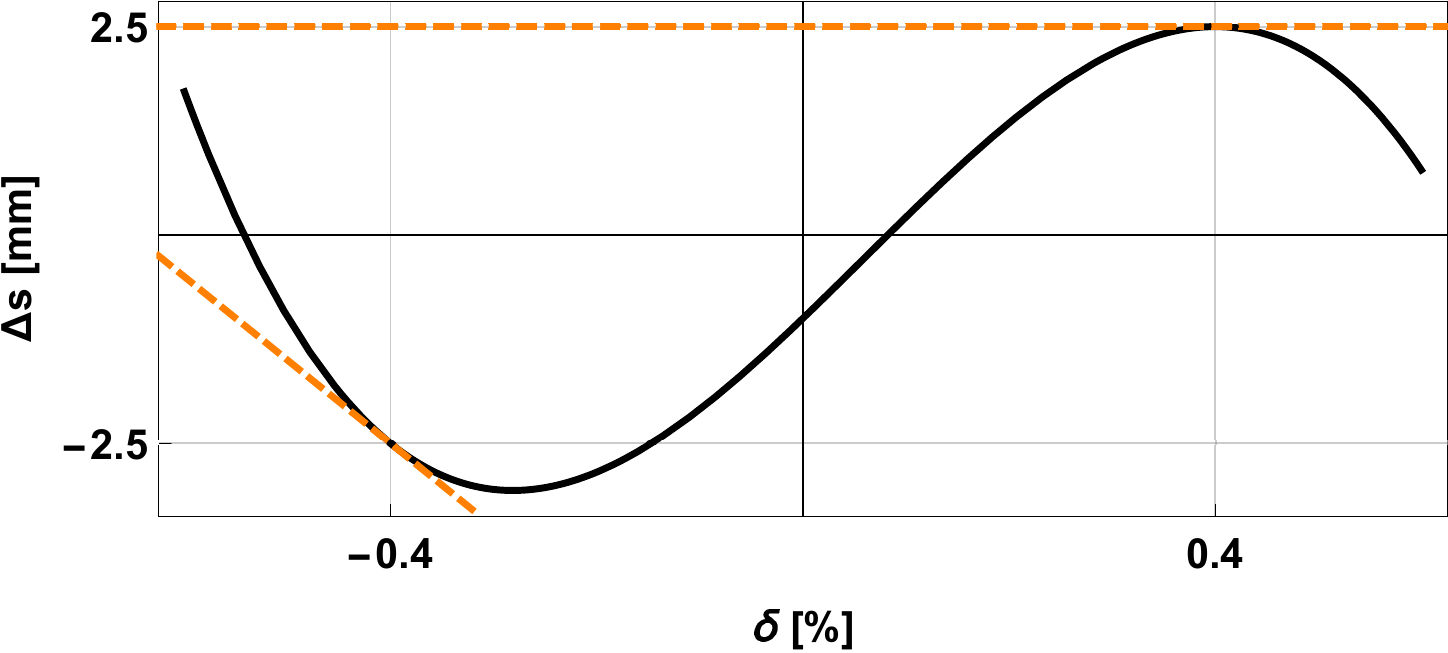}
    \caption{Arc path length as a function of relative momentum deviation where we choose reference momentum, first, second and third order longitudinal dispersions. This allows independent control over path length and $R_{56}$ of two off-momentum beams. Example shows for two beams at $\delta = \pm 0.4\%$ the path length difference is $5$~mm with effective $R_{56} = 0$~mm and $-10$~mm respectively, illustrated by the dashed orange lines.}
    \label{fig:CTpathlength}
\end{figure}

The idea of exploiting the different beam energies accelerating and decelerating within the same arc can be extended to independently control the path lengths and linear longitudinal dispersions of both accelerating and decelerating beams with the right choice of on-momentum first, second and third order longitudinal dispersions. This added flexibility to common transport arcs would however require sextupoles and octupoles to adequately set the higher order longitudinal dispersions while still keeping control over the transverse dispersions and chromatic amplitudes with a wide energy acceptance. Since this method does not provide control over the higher order dispersions as seen by the off-momentum beams, it is potentially useful to implement in arcs where the beam chirp is expected to be zero, and recuperate some path length control between accelerating and decelerating arcs. Figure~\ref{fig:CTpathlength} shows an example where the higher energy beam has a path length \SI{5}{mm} longer than the lower energy beam and their effective $R_{56}$ values are \SI{0}{mm} and \SI{-10}{mm} respectively. The on-momentum longitudinal dispersions $R_{56}$, $T_{566}$ and $U_{5666}$ are \SI{1.21}{m}, \SI{-24.2}{m} and \SI{-35156}{m} respectively.

\begin{figure}[!ht]
    \centering
 \includegraphics[width=0.48\textwidth]{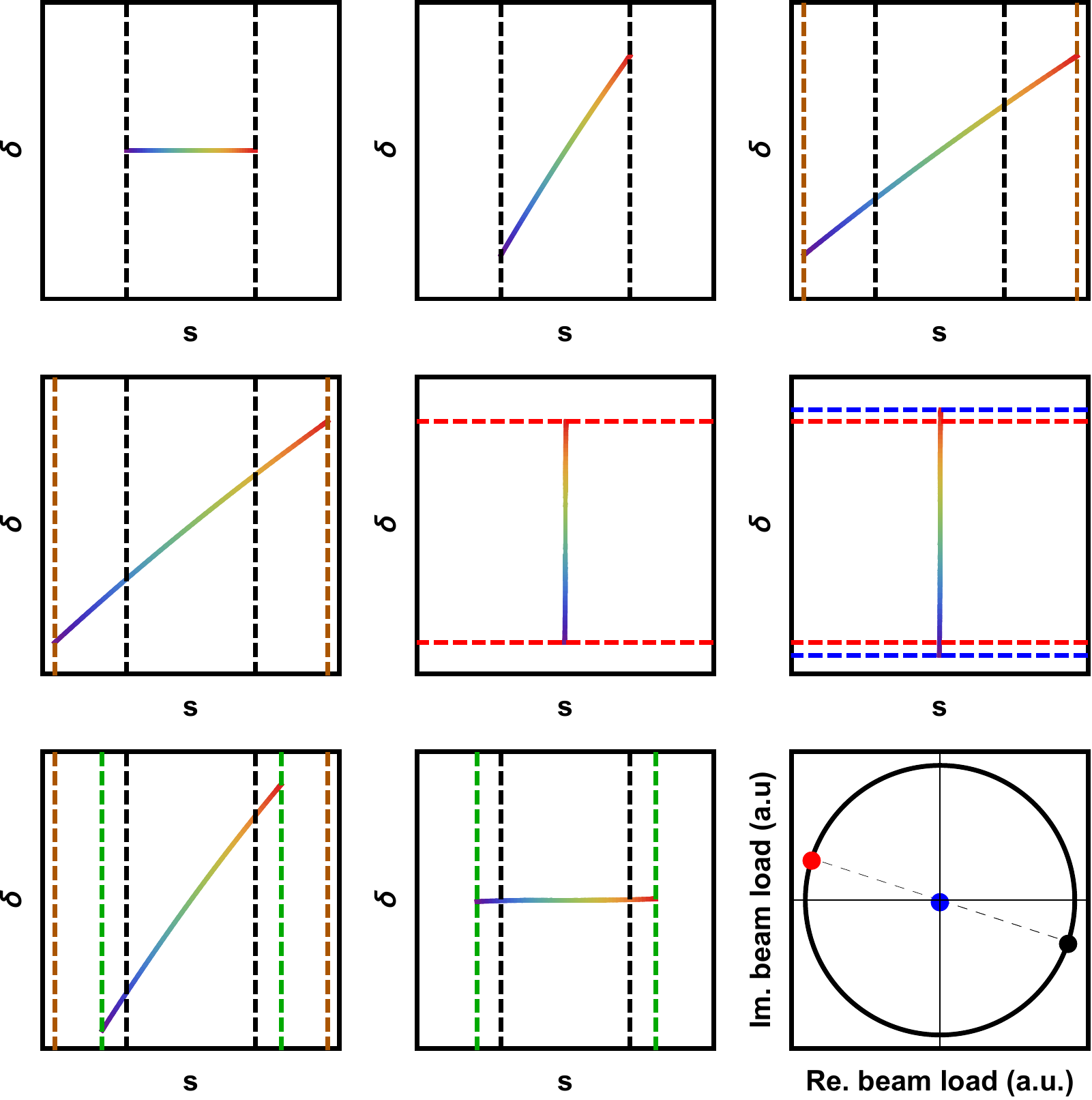}
    \caption{Equivalent of Example IA shown in Fig.~\ref{fig:compressSR} but bunch decompression does not occur at top energy. The bunch undergoes decompressions during acceleration and deceleration resulting in a bunch longer at the dump than at the injector. Black dashed lines indicate initial bunch length, orange dashed line indicates maximum bunch length, green dashed line indicates final bunch length, red dashed line indicates bunch energy spread at top energy and blue dashed line indicates adiabatic growth of energy spread of the decelerating fully compressed bunch.}
    \label{fig:ER@CEBAFmatch}
\end{figure}

We also consider a configuration where the top energy arc can only be used to compress the bunch, but not to decompress it, as shown in Fig.~\ref{fig:ER@CEBAFmatch}. In this case, we can set an intermediate arc $R_{56}$ to have the opposite sign. With this scheme, the bunch decompression (black and orange dashed lines) is larger during deceleration (green dashed line) thanks to the combination of the energy spread growth in the interaction region and the adiabatic growth of the relative energy spread in the decelerating RF phases between the top and the decompressing arc (red and blue dashed lines).

\section{Parasitic crossings}
Parasitic crossings, also know as overcompressions, where the bunch head and tail exchange places, provide an additional tool to find longitudinal matches in ERLs as, in effect, they allow the sign of the beam chirp during transport to change between linac passes. Within our model, this corresponds to a negative inverse global compression function $Z_{i}<0$. One could expect significant degradation to occur at a parasitic compression, and this would be of particular concern during acceleration. However if the minimum bunch length during this compression is relatively large due to the presence of uncompensated RF curvature at that location, such degradation would be not significant. One can picture this as the bunch ``rolling" through a ``banana" shape in the phase space, as opposed to standing totally upright.\par
\begin{figure}[!ht]
    \centering
    \includegraphics[width=0.48\textwidth]{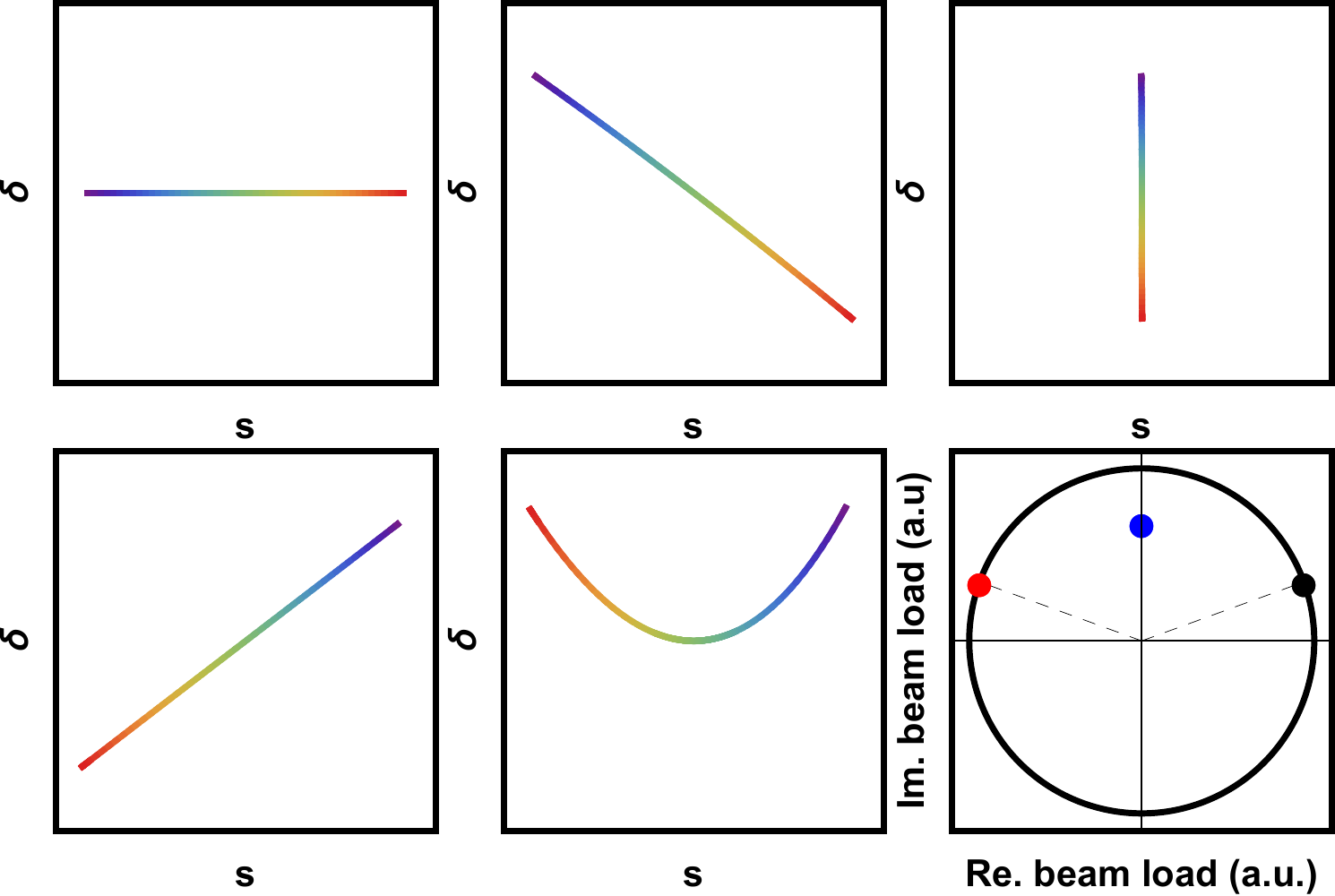}
    \caption{Equivalent of Example IA shown in Fig.~\ref{fig:compressSR} but with utilisation of a parasitic crossing and deceleration on opposite side of rf trough in order to remove linear chirp. This results in a large imaginary resultant RF load. We see residual curvature as the natural $T_{566}$ value of the arcs add to the RF curvature for deceleration on falling side of trough.}
    \label{fig:CompressiveParasiticTopArc}
\end{figure}

Bunch decompression immediately after the IP in a compressive match can be such that the bunch chirp is of the opposite sign before and after. The bunch undergoes a parasitic crossing and the bunch energy spread can still be compressed during deceleration by changing the side of trough it is decelerated on with the effect of an imaginary resultant beam load. The collective effects during this process will degrade the beam's emittance, however this happens after the interaction region and control of higher order longitudinal dispersion can be used to ensure energy recovery remains satisfied. This sequence of longitudinal manipulations is shown in Fig.~\ref{fig:CompressiveParasiticTopArc}.
\par
\begin{figure}[!htb]
    \centering
    \includegraphics[width=0.5\textwidth]{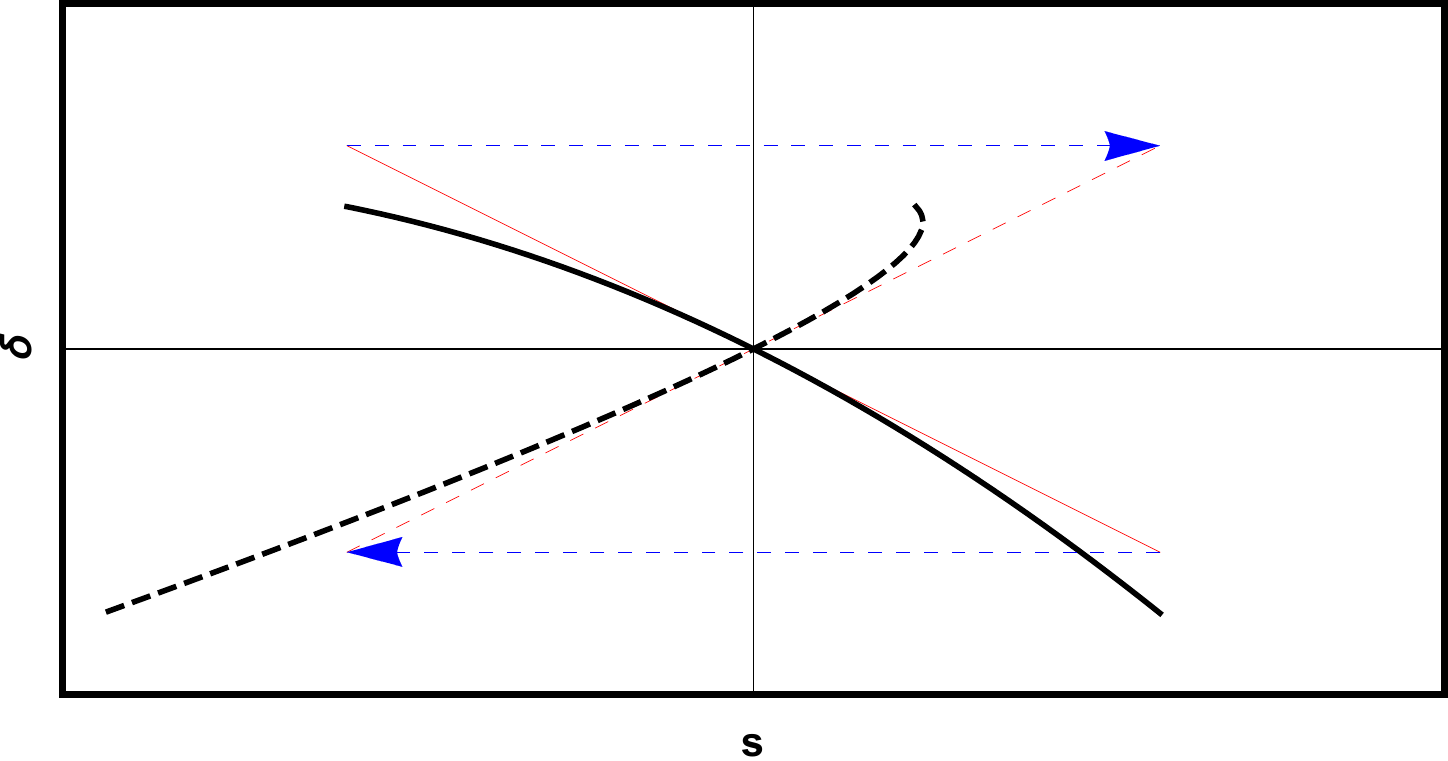}
    \caption{Change in curvature as an example bunch (black) undergoes a linear compression (solid to dashed). The same compression acting on a linearized bunch is shown in red.}
    \label{fig:ShearCrossing}
\end{figure}

 First-order transformations like these, via control of $R_{56}$, can be used to control second-order parameters of our beam and completely or partially cancel the effects of the rf curvature, as shown in Fig.~\ref{fig:ShearCrossing}, such that it is then compensated by the remaining RF passes. In order to continue the chirp compensation, the subsequent passes must be on the opposite side of the waveform.\par
\begin{figure*}[!htb]
    \centering
    \includegraphics[width=\textwidth]{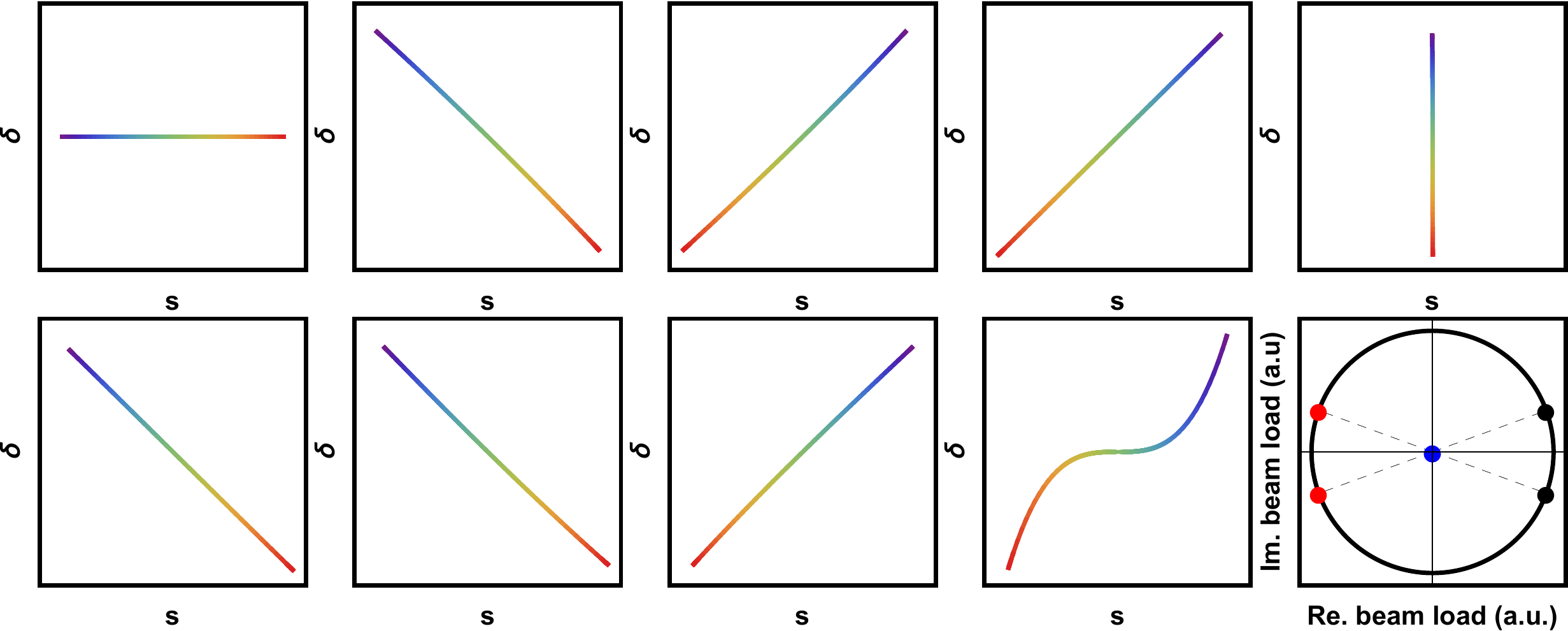}
    \caption{Longitudinal match solutions using parasitic crossings to cancel rf curvature in common transport. RF beam load plot (bottom right) shows phase choices during acceleration (black), deceleration (red), and resultant (blue).}
    \label{fig:CompressiveParasiticASMLcommon}
\end{figure*}

This mechanism, if implemented in a common transport configuration with shared longitudinal dispersions and arc path-lengths requires parasitic crossings during acceleration, deceleration and at top energy, as shown in Fig.~\ref{fig:CompressiveParasiticASMLcommon}. The parasitic crossing during acceleration would have the same linearizing effect as during deceleration, but it would also increase the beam emittance before the interaction point. This alternative method of linearization is also applicable in energy spread minimizing matches with the exception of crossings happening at the top energy since the beam would have zero chirp at that point.\par
\begin{figure*}[!htb]
    \centering
    \includegraphics[width=\textwidth]{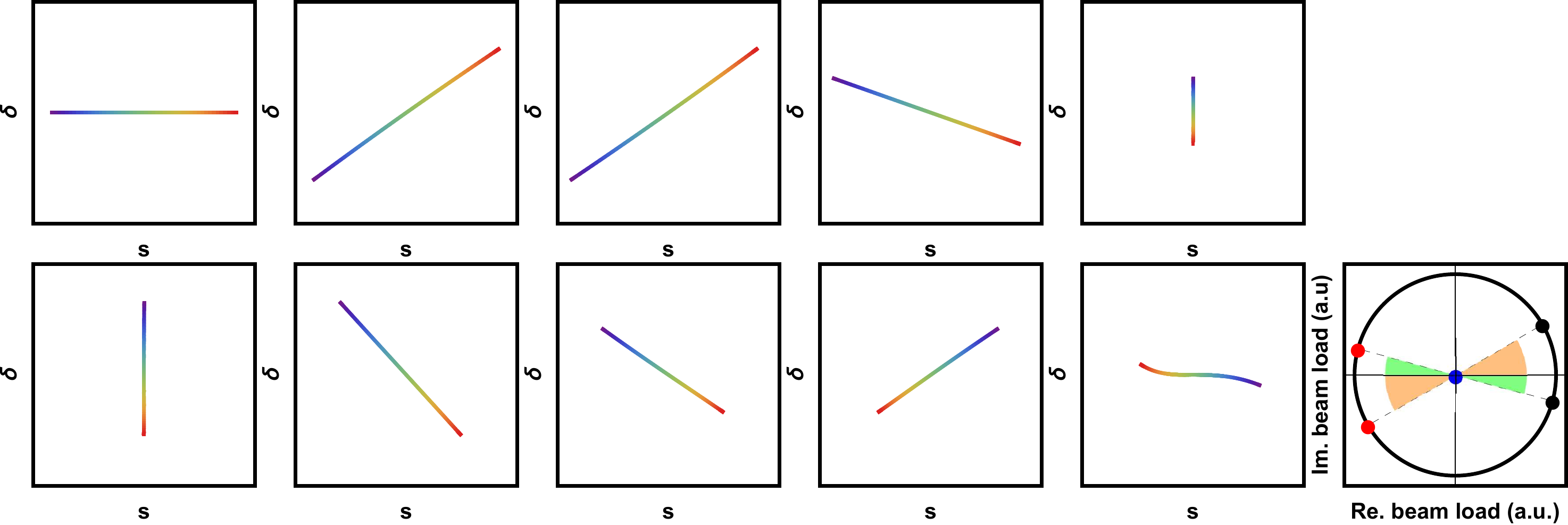}
    \caption{Compressive longitudinal match with energy spread growth at the interaction point. Longitudinal phase spaces during acceleration (top) and during deceleration (bottom) with a parasitic crossing. RF beam load plot (bottom right) shows phase choices during acceleration (black), deceleration (red), and resultant (blue) with highlighted angles representing matching magnitudes.}
    \label{fig:FELmatch}
\end{figure*}

The separate transport configurations' independent control over each arc's longitudinal dispersions enables the use of this transformations during deceleration without compromising the beam quality before it reaches the interaction region. As the beam quality constraints during deceleration are relaxed, a separate transport can also compress the bunch during deceleration to cope with the energy spread increases expected from an FEL interaction. This is showcased in Fig.~\ref{fig:FELmatch}, showing the beam during acceleration without compressions or parasitic crossings and during deceleration after doubling its energy spread.
\section{Conclusions}
In this paper we have shown possible longitudinal matches for a wide range of multi-pass ERL configurations comprising  compressive matches and energy spread minimising matches for common transport and separate transport topologies and with and without synchrotron radiation compensation as summarized by Fig.\ref{fig:treeBW}. 
We conclude that for systems with negligible energy losses, arc path length and longitudinal dispersion configurations exist for both compressive matches and energy spread minimising matches for both common transport and separate transport topologies, although common transport matches will require a more intricate linearizaiton scheme to obtain linearized bunches at both the interaction point and dump.\par
If synchrotron radiation energy losses must be compensated, we show solutions for compressive matches and energy spread minimising matches for separate transport configurations. Synchrotron radiation compensating compressive matches are also available in common transport configurations. However, synchrotron radiation compensating energy spread minimising matches in a common transport configuration require transport between rf passes with energy acceptances of a few \% as shown in Fig.~\ref{fig:LHeCArc1Acceptance} or require strong bunch length modulations including parasitic crossings throughout all of the transport, especially if peak energies are in the range of $\gtrsim\SI{50}{GeV}$ as proposed for LHeC.\par
Throughout this analysis, no collective effects have been taken into account. The two collective effects that will have the highest impact on the longitudinal phase space will be coherent synchrotron radiation (CSR) and longitudinal space charge (LSC). CSR will lower the energy at the center of the bunch with respect to the tails. This is opposite to the curvature imprinted by the rf on an accelerating beam and thus will reduce our linearization requirements. However, during deceleration, the changes in the longitudinal phase space from CSR will add to the decelerating RF curvature which together with the adiabatic growth of the energy spread during deceleration will result in a significant energy spread at the dump if not accounted for. LSC can be introduced in our considerations by tracking the bunch through the low energy sections~\cite{Khan2019a} and taking the pre-accelerated bunch as the start of our analysis and tracking the last decelerating pass towards the dump.\par
With these caveats, we have demonstrated a methodology for designing multi-pass ERLs for a wide range of applications.

\section{Acknowledgments}
This work was partially funded by STFC grant ST/P002056/1 ``The Cockcroft Institute of Accelerator Science and Technology".

\bibliographystyle{apsrev}
\bibliography{references}
\end{document}